\documentclass[%
 reprint,
 amsmath,amssymb,
 aps,longbibliography,
superscriptaddress
]{revtex4-2}

\usepackage[utf8]{inputenc}
\usepackage{graphicx}% Include figure files
\usepackage{bm}% bold math
\usepackage{bbold}
\usepackage{braket}
\usepackage{tikz,pgfplots}
\usepackage{tikz-qtree}
\usepackage{xcolor}
\usepackage{tabularx}
\usepackage{booktabs}
\usepackage[outdir=./]{epstopdf}
\pgfplotsset{compat=1.10}
\usepackage{siunitx}

\usetikzlibrary{patterns,shadows,fadings,positioning,trees,calc}
\usetikzlibrary{shapes}

\DeclareMathOperator\erf{erf}

\tikzfading[name=fade cloud,
         inner color=transparent!99,
         outer color=transparent!40]
\tikzfading[name=fade inside,
         inner color=transparent!80,
         outer color=transparent!30]
\tikzfading[name=fade out,
         inner color=transparent!0,
         outer color=transparent!90]
\tikzfading[name=fade inner,
         inner color=transparent!100,
         outer color=transparent!0]

\definecolor{diplom1}{RGB}{101 156 239}
\definecolor{diplom2}{RGB}{000 000 128}
\definecolor{diplom3}{RGB}{153,0,0} %unirot
\definecolor{diplom4}{RGB}{232,215,23}
\definecolor{diplom5}{RGB}{51,37,60}

\definecolor{unirot}{RGB}{153,0,0}
\definecolor{unirot_hell}{RGB}{255,228,225}
\definecolor{lightblue}{RGB}{242.2,249.88,255}

\AtBeginDocument{
\heavyrulewidth=.08em
\lightrulewidth=.05em
\cmidrulewidth=.03em
\belowrulesep=.65ex
\belowbottomsep=0pt
\aboverulesep=.4ex
\abovetopsep=0pt
\cmidrulesep=\doublerulesep
\cmidrulekern=.5em
\defaultaddspace=.5em
}

\begin{document}

%\preprint{APS/123-QED}

\title{Electronic decay process spectra including nuclear degrees of freedom}

\author{Alexander V.\ Riegel}%
\affiliation{%
Center for Light-Matter Interaction, Sensors \& Analytics (LISA$^+$),
University of Tübingen\\
Auf der Morgenstelle 15, 72076 Tübingen, Germany
}%
\affiliation{%
Institute of Physical and Theoretical Chemistry,
University of Tübingen\\
Auf der Morgenstelle 18, 72076 Tübingen, Germany
}%
\author{Elke Fasshauer}
 \email{elke.fasshauer@gmail.com}
\affiliation{%
Center for Light-Matter Interaction, Sensors \& Analytics (LISA$^+$),
University of Tübingen\\
Auf der Morgenstelle 15, 72076 Tübingen, Germany
}%
\affiliation{%
Institute of Physical and Theoretical Chemistry,
University of Tübingen\\
Auf der Morgenstelle 18, 72076 Tübingen, Germany
}%
\affiliation{
Department of Chemistry, University of Oxford, Oxford OX1 3QZ, United Kingdom
}

\date{\today}% It is always \today, today,

\begin{abstract}
In the field of chemistry, where nuclear motion has traditionally been a focal point,
we now explore the ultra-rapid electronic motion spanning attoseconds to femtoseconds,
demonstrating that it is equally integral and relevant to the discipline.
The advent of ultrashort attosecond pulse technology has revolutionized our ability
to directly observe electronic rearrangements in atoms and molecules,
offering a time-resolved insight into these swift processes. Prominent examples
include Auger-Meitner decay and Interparticle Coulombic Decay (ICD).
However, the real challenge lies in interpreting these observations,
where theoretical models are indispensable.

Building upon the analytical framework introduced in Phys.\ Rev.\ A 101, 043414 (2020),
which analyzed the spectra of electrons emitted during electronic decay processes
from a purely electronic perspective, our paper represents a significant advancement.
We extend this theoretical base to include nuclear dynamics, utilizing the
Born-Oppenheimer approximation to deepen our understanding of the intricate
interaction between electronic and nuclear motion in these processes.

We illustrate the impact of incorporating nuclear degrees of freedom in several
theoretical cases characterized by different numbers of vibrational bound states
in both the electronic resonance and the electronic final state. This approach
not only clarifies complex spectral features and unusual peak shapes but also
demonstrates a method for extracting the energy differences between multiple
vibrational resonance states through their distinctive interference patterns.
\end{abstract}

\maketitle

%%%%%%%%%%%%%%%%%%%%%%%%%%%%%%%%%%%%%%%%%%%%%%%%%%%%%%%%%%%%%%%%%%%%%%%%%%%%%%%%%
%\input{subsecs/intro}

\section{Introduction}
\label{sec:intro}

Since the advent of ultrashort laser pulses
in the attosecond and femtosecond range \cite{Orfanos19},
it has in principle become possible to study the time evolution of chemical
reactions in detail.
However, the duration of a laser pulse is inversely proportional to its
energetic width and this width, for the current state of the art,
needs to include at least one
half-cycle of the mean photon energy. Hence, ultrashort laser pulses
typically have mean photon energies in the extreme ultraviolet (XUV) or even
x-ray range. We therefore aim to understand the processes initiated by these
ultrashort laser pulses and what information we can extract from their respective
time-resolved spectra.

The interaction of ultrashort laser pulses
with matter prompts ionizations and excitations from the inner
shells of atoms and molecules. The resulting excited states can then decay under
electronic rearrangement and emission of an electron, which carries the excess
energy.
Such electronic decay processes have been known for many years. The
Auger-Meitner process was discovered about 100 years ago \cite{Meitner22,Auger23}
and also the first theoretical prediction of the Interparticle Coulombic Decay (ICD)
is more than two decades old \cite{Cederbaum97}.

The different electronic decay processes can be distinguished by the number of involved
entities (atoms, molecules or weakly bound conglomerates) and the
kind of electronic rearrangement during the process.
In the Auger-Meitner process, only one entity $A$ is involved. After primary
ionization from an inner shell, the vacancy is filled by an electron from an
outer shell and another electron is emitted:

\begin{equation*}
 A \xrightarrow{h\nu} A^{**+} + e_\text{photo}^- \rightarrow A^{2+} + e_\text{photo}^- + e_\text{Auger-Meitner}^- \, .
\end{equation*}
Here, we indicate an inner-shell vacancy by two stars $^{**}$.

After an inner-shell excitation, the excited electron can either fill the vacancy
itself and cause the emission of another electron (participator resonant
Auger-Meitner process)
or it can stay excited while another electron fills the vacancy and a third
electron is emitted (spectator resonant Auger-Meitner process):

\begin{alignat*}{5}
 &A &&\xrightarrow{h\nu} A^{**}
    &&\rightarrow A^{+} &&+ e_\text{Auger-Meitner}^- \quad &&\text{(participator)}\\
 &A &&\xrightarrow{h\nu} A^{**}
    &&\rightarrow A^{*+} &&+ e_\text{Auger-Meitner}^- \quad  &&\text{(spectator)} .
\end{alignat*}
After the spectator Auger-Meitner process, the system is necessarily still 
excited and will decay further, whereas
for the participator Auger-Meitner process with only
one decay channel, this is not the case.
However, when more channels are accessible, it
is also possible to create an excited state, which will
decay further.

The Auger-Meitner process is found in a variety of different systems
like atoms, molecules and solids.
As a consequence and because it is element specific, Auger-Meitner electron spectroscopy
is used for surface analysis in metallurgy,
quality analysis of microelectronics
as well as for basic studies of chemical reaction mechanisms \cite{AES_Seah_86}.
It has also been a test process for
time-resolved measurements in general
{\cite{Drescher02,Smirnova03,Babushkin22}}
and is often observed as a side product of modern x-ray spectroscopies
\cite{Greczynski20}.

In the ICD process, the energy released during the vacancy filling is transferred through
space to a different entity, which consequently emits the ICD electron:
\begin{equation*}
 AB \xrightarrow{h\nu} A^{**+}B + e_\text{photo}^- \rightarrow A^{+} + B^+ + e_\text{photo}^- + e_\text{ICD}^- \, .
\end{equation*}
After the process, the two involved entities are both positively charged and therefore
repel each other. The ICD process also has its resonant variants (resonant ICD, RICD),
in analogy to the Auger-Meitner process.

ICD and ICD-like processes occur in many different systems with weakly
interacting units such as noble gas clusters
\cite{Santra00_1,Hergenhahn11,Jahnke15,Fasshauer16},
proteins \cite{Harbach13}, solvents
\cite{Mueller06,Hergenhahn11,Jahnke15,Richter18,
Kryzhevoi11_1,Stoychev11,Kryzhevoi11_2,Oostenrijk18}
as well as semi-conductors
\cite{Bande11,Bande13b,Dolbundalchok16}
and are discussed as a key mechanism of radiative DNA damage
\cite{Alizadeh15,Boudaiffa00,Martin04,Brun09,Surdutovich12}.

In this article, we will focus on the resonant electronic decay processes, which are
initiated by inner-shell excitation rather than by ionization.

For time-resolved spectroscopy of electronic decay processes, most of the developed
theory considers atomic systems \cite{Wickenhauser05,Wickenhauser06},
for whose description the purely electronic
Schrödinger equation suffices. However, for Auger-Meitner processes of molecules,
molecules bound to a surface, like in heterogeneous catalysis, or an ICD process,
the nuclear degrees of freedom will affect the spectra of the emitted electrons.
Since numerical solutions of both the electronic and the nuclear dynamics are
expensive to solve numerically, we aim for an analytical description following
the basic formulation of Ref.~\cite{Fasshauer20_1}, from which
we can draw conclusions about basic properties. Moreover, by careful
observation of the assumptions we make in our derivations, it will be possible
to assign further properties observed in experiment to those parts which we
neglect.

The analytical description of time-independent spectra of electronic decay processes
dates back to the work by Fano \cite{Fano61}. His basic theory focuses on
purely electronic wavefunctions. It has been generalized to quantum defect
theory for molecules as well \cite{Greene79,Greene82,Greene_erratum_84,Greene85,bk:fano1986}.
The latter has been investigated in a time-dependent approach for rotationally
autoionizing states of the hydrogen molecule \cite{Kirrander07}.
We will provide a similar
ansatz for the wavefunction of a decaying state including both electronic and
nuclear eigenstates of the system for non-interacting resonance states within the
Born-Oppenheimer approximation, show
general properties of the time-resolved spectra for selected cases and
discuss the limits of our approach based on the approximations we introduce.

The article is structured as follows:
In section \ref{sec:theory}, we choose an ansatz for the Fano wavefunction including
nuclear degrees of freedom for one electronic resonance and one electronic final state
and derive the corresponding coefficients.
We then derive
the working expressions for the simulation and analysis of the time-resolved
kinetic energy spectra of the emitted electron. In section \ref{sec:comp}, we
have collected all the general parameters which we use to simulate the
spectra for different showcases presented in section \ref{sec:cases}.
We discuss the limitations and consequences of our
introduced approximations in section \ref{sec:approx} and summarize in
section \ref{sec:summary}.

\section{Theory}
\label{sec:theory}

%In this section, we derive the explicit time- and energy-differential ionization
%probability formulas for an RICD process. We furthermore sketch the origin of
%Eqs.~(\ref{eq:gauss_lorentz}) and (\ref{eq:oscillation}).

\begin{figure}[!ht]
 \centering
 \includegraphics[width=0.79\columnwidth]{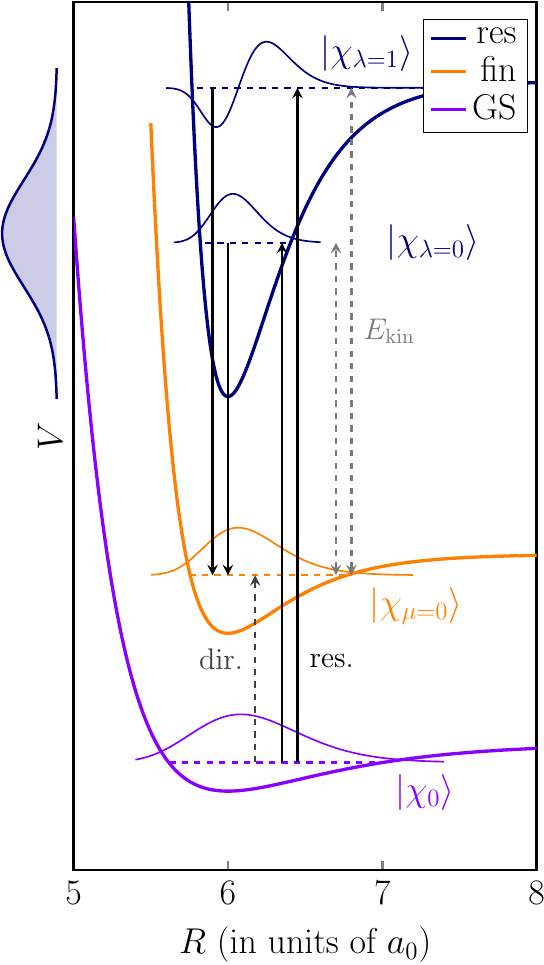}
 \caption{(Colour online) Illustration
 \cite{figshare_pics_2024}
 of the energy levels in the model systems used in this article.
 The three potential curves belong to the electronic ground state (lower curve in violet),
 final state (middle curve in orange) and resonance state (upper curve in dark blue).
 For each potential, the position of the bound vibrational eigenstates
 is indicated by dashed lines (here: one in both the electronic ground state and the final state,
 two in the resonance state, which corresponds to case 4 in section \ref{sec:2l1m})
 together with the corresponding vibrational eigenfunction and our chosen symbolic designation.
 The one-headed arrows depict possible pathways
 for the transition from the electronic ground state to the final state, either directly
 (dashed arrow) or via one of the vibrational eigenstates of the electronic resonance state
 (solid arrows). The double-headed arrow labelled "$E_\mathrm{kin}$"
 corresponds to the kinetic energy of the RICD electron emitted
 following the transition from the resonance state to the final state,
 in this case between the vibrational ground states.
 The Gaussian function on the left represents the initiating ultrashort laser pulse,
 which is energetically centred around the energy of the vibrational ground state
 of the electronic resonance state, relative to the electronic ground state.
 }
 \label{fig:energy_overview}
\end{figure}

The electronic decay processes we intend to describe fully time-dependently
are schematically illustrated in Figure~\ref{fig:energy_overview} for the specific case
of two nuclear bound states in the resonance state and one nuclear bound state in the
electronic final state. The aim is to find a description for any combination
of nuclear bound states in the electronic resonance and final states.

Starting from the ground state $\ket{G}$, we excite the system with an XUV pulse with
a mean photon energy $\Omega$ and thereby create a coherent superposition
of the two nuclear states $\ket{\chi_\lambda}$
of the electronic
resonance state $\ket{r}$. The minimum of the potential energy curve is $E_r$ and
the energies of the nuclear states $E_\lambda$ are determined relative to $E_r$.
These resonance states can then, under emission of an electron,
decay to a continuum state characterized by an electronic final state
$\ket{E}$ and a nuclear final state $\ket{\chi_\mu}$ with a
lifetime $\tau_\lambda$.
The continuum states related to spectator resonant electronic decay processes
can couple to other continuum states under emission
of a photon. These processes are usually several orders of magnitude slower
than the electronic
decay process and are therefore neglected in this work.
Alternatively to arriving in the continuum via a decaying resonance state, a simultaneous
direct excitation and ionization (spectator) or a direct ionization (participator)
channel is also possible, which is denoted "dir" in the figure.

From the derivations of Ref.~\cite{Fasshauer20_1}, we know that the time-independent
Fano wavefunction is essential to the description of the time-resolved spectra.
We therefore first discuss the derivation of approximative Fano wavefunctions
including nuclear degrees of freedom and afterwards derive the expression for
the time- and energy-differential ionization probability.

%%%%%%%%%%%%%%%%%%%%%%%%%%%%%%%%%%%%%%%%%%%%%%%%%%%%%%%%%%%%%%%%%%%%%%%%%%%%%%%%%
%\input{subsecs/Fano}

\subsection{Derivation of the Fano coefficients including nuclear wavefunctions}
\label{app:Fano_coeff}
Fano's purely electronic and time-independent wavefunction
for one resonance and one final state is \cite{Fano61}

\begin{equation}
\ket{\Psi_{\underline{E}}} = 
                                a(\underline{E}) \ket{r}
                              + \int \mathrm{d}E'
                                b_{E'}(\underline{E}) \ket{E'} .
\end{equation}
Here, $\ket{r}$ is the electronic wavefunction of the bound resonance state and
$\ket{E'}$ is the electronic continuum wavefunction including both
the bound final state and the emitted free electron. Both are considered to
be eigenstates of the Hamiltonian encompassing the operator of the
unperturbed system as well as the configuration interaction operator.
The latter is mediating the electron-electron interaction
which is causing the decay.
$a$ and $b$ are coefficients, which
are determined for the different cases.

We extend this ansatz to include nuclear degrees of freedom in the Born-Huang
expansion \cite{bk:bornhuang1954}
by adding a summation over the different nuclear eigenstates
of the two respective potential energy curves. Hereby, we assume
the electronic
wavefunctions to be adiabatic wavefunctions.
\begin{equation}
 \ket{\Psi_{\underline{E}}} = \sum\limits_\lambda
                                a_\lambda(\underline{E}) \ket{r} \ket{\chi_\lambda}
                              + \sum\limits_{\mu'} \int \mathrm{d}E'
                                b_{E',\mu'}(\underline{E}) \ket{E'} \ket{\chi_{\mu'}}
\end{equation}

In order to determine the coefficients, we follow the same strategy as for the
pure electronic Fano ansatz in Ref.~\cite{Fano61} and solve the following set
of equations assuming the Born-Oppenheimer approximation
\cite{Born27,Ballhausen72}.

\begin{align}
 \braket{r| \hat{H} |\Psi_{\underline{E}}}
  =& \sum\limits_\lambda \braket{r| \hat{H} a_\lambda(\underline{E}) |r}
     \ket{\chi_\lambda}    \nonumber \\
   &+ \sum\limits_{\mu'} \int \mathrm{d}E'
     \braket{r| \hat{H} b_{E',\mu'}(\underline{E}) |E'} \ket{\chi_{\mu'}} \nonumber \\
  =& \, \underline{E} \braket{r | \Psi_{\underline{E}}}
\end{align}
\begin{align}
 \braket{E''| \hat{H} |\Psi_{\underline{E}}}
  =& \sum\limits_\lambda \braket{E''| \hat{H} a_\lambda(\underline{E}) |r}
     \ket{\chi_\lambda}    \nonumber \\
   &+ \sum\limits_{\mu'} \int \mathrm{d}E'
     \braket{E''| \hat{H} b_{E',\mu'}(\underline{E}) |E'} \ket{\chi_{\mu'}} \nonumber \\
  =& \, \underline{E} \braket{E'' | \Psi_{\underline{E}}}
\end{align}
\normalsize

Here, the Hamilton operator, as in Fano's ansatz,
consists of the unperturbed Hamiltonian of the system,
$H_0$, which involves both electronic and nuclear operators and
is diagonal in the chosen electronic basis,
and the purely electronic
so-called configuration interaction operator $V$
coupling the electronic resonance and continuum eigenstates with each other:
\begin{equation} \label{eq:Fano_Hamiltonian}
H_F = H_0 + V     \, .
\end{equation}

The vibronic eigenfunctions $\left\{ \ket{r}\ket{\chi_\lambda} \right\}$
and $\left\{ \ket{E}\ket{\chi_\mu} \right\}$ of $H_0$ are characterized by
their respective energies, which we choose to give as the sum of the minimum
energy of the electronic potential energy curve plus the energy of the
energy of the different vibrational levels ($E_r,\text{min} + E_\lambda$
and $E_\text{fin,min} + E_\mu$). For a shorter notation, we will
from here on denote them as $E_r$ and $E_\text{fin}$.
We assume that the coefficients are not explicitly dependent on either the
electronic or the nuclear coordinates to obtain:

\begin{align}
 \braket{r| \hat{H}_F |\Psi_{\underline{E}}}
  =& \sum\limits_\lambda  a_\lambda(\underline{E}) (E_r + E_\lambda)
     \ket{\chi_\lambda}  \nonumber \\
  &+ \sum\limits_{\mu'} \int \mathrm{d}E'
     b_{E',\mu'}(\underline{E}) V_{rE'} \ket{\chi_{\mu'}} .
\end{align}

We project out the contribution of a single nuclear eigenfunction
 $\bra{\chi_{\lambda'}}$ and in doing so assume that the electronic
configuration interaction term $V_{rE'}$
 does not directly depend on the nuclear coordinates.
We have then renamed $\lambda' \rightarrow \lambda$ :
\begin{align}
 \bra{\chi_\lambda} \braket{r| \hat{H}_F |\Psi_{\underline{E}}}
  =& a_\lambda(\underline{E}) (E_r + E_\lambda)  \nonumber \\
   &+ \sum\limits_{\mu'} \int \mathrm{d}E'
     b_{E',\mu'}(\underline{E}) V_{rE'} \braket{\chi_\lambda |\chi_{\mu'}} \nonumber \\
  =& \underline{E} \, a_\lambda(\underline{E})  \, . \label{eq:a}
\end{align}

Analogously, we obtain
\begin{align}
	\bra{\chi_{\mu'}} \braket{E'| \hat{H}_F |\Psi_{\underline{E}}}
	=& \sum\limits_\lambda a_\lambda(\underline{E}) V_{E'r}
	\braket{\chi_{\mu'} | \chi_\lambda} \nonumber \\
	&+ 
	b_{E',{\mu'}}(\underline{E}) (E' + E_{\mu'}) \nonumber \\
	=& \underline{E} \, b_{E',\mu'}(\underline{E})     \, .	\label{eq:b}
\end{align}

By solving Eq.~(\ref{eq:b}) for $b_{E',\mu'}(\underline{E})$
we arrive at
\begin{align}
  b_{E',\mu'}(\underline{E})
=& \sum\limits_\lambda
  \frac{a_\lambda(\underline{E}) V_{E'r} \braket{\chi_{\mu'} | \chi_\lambda}}
  {\underline{E} - E' - E_{\mu'}}   \nonumber  \\
&+ \sum\limits_\lambda a_\lambda(\underline{E}) V_{E'r} \braket{\chi_{\mu'} | \chi_\lambda}
  z(\underline{E}) \delta(\underline{E} - E' - E_{\mu'})    \, .
\end{align}

Here, we implicitly assume that the first part of the sum becomes the principal
part upon integration over one of the energies.
For only one resonance and one final state,
we would insert this expression into Eq.~(\ref{eq:a})
and solve for $z(\underline{E})$. However, when including nuclear wavefunctions, we
arrive at

\begin{widetext}
\begin{equation} \label{eq:Fano_approx}
 (\underline{E} - E_r - E_\lambda) \, a_\lambda(\underline{E})
 = \sum\limits_{\mu',\lambda'} \mathcal{P} \int\limits \mathrm{d} E' \,
   \frac{a_{\lambda'}(\underline{E}) |V_{E'r}|^2
     \braket{\chi_\lambda | \chi_{\mu'}} \braket{\chi_{\mu'} | \chi_{\lambda'}} }
   {\underline{E} - E' - E_{\mu'}}
 + \sum\limits_{\mu',\lambda'}
   a_{\lambda'}(\underline{E}) |V_{(\underline{E}-E_{\mu'})r}|^2
   z(\underline{E})
     \braket{\chi_\lambda | \chi_{\mu'}} \braket{\chi_{\mu'} | \chi_{\lambda'}} .
\end{equation}
\end{widetext}

An analytical closed form solution of this equation is not possible because only the decay of several
resonance states to the same final state or the decay of one resonance state into
several final states can be properly normalized in the way
discussed by Fano \cite{Fano61}. In our case,
the sum over $\mu'$ or $\lambda'$ prevents this normalization. We therefore make the
assumption that the nuclear wavefunctions of the resonance state are not coupled
via the continuum and hence introduce $\delta_{\lambda,\lambda'}$.
This approximation is valid when the coupling of different nuclear resonance states
is small or
when, for a given $\mu'$, the Franck-Condon factor for $\lambda$
is significantly larger
is significantly larger than for all other values
of $\lambda'$. Then, all contributions other than the one for $\lambda'=\lambda$
are small and a large part of the sum is covered by the main contribution.
We therefore limit the scope of this manuscript to so-called
non-overlapping vibronic resonance states \cite{mies1968}.

With this in mind, we deduce
\begin{equation} \label{eq:z_nucl}
 z(\underline{E}) =
 \frac{\underline{E} - E_r - E_\lambda - \sum\limits_{\mu'}F_{\mu'}(\underline{E})}
 {W_\lambda}    \, .
\end{equation}

Here, $W_\lambda$ is the weighted sum of absolute squares of the
configuration interaction terms of the different nuclear final states for a single
nuclear resonance state
\begin{equation} \label{eq:Wlambda}
W_\lambda
= \sum_{\mu'} |V_{(\underline{E}-E_{\mu'})r}|^2
  | \braket{\chi_{\lambda}|\chi_{\mu'}} |^2
\end{equation}
and
$F_{\mu'}(\underline{E}) = \mathcal{P} \int\limits \mathrm{d} E' \,
   \frac{ |V_{E'r}|^2 |\braket{\chi_\lambda|\chi_{\mu'}}|^2 }
   {\underline{E} - E' - E_{\mu'}}$ .

Here, we have assumed that the electronic coupling matrix elements
$V_{(\underline{E}-E_{\mu'})r}$ are independent of the interparticle distance $R$. In reality, they exhibit a structure dependence, e.g.,
an exponential behaviour for Auger-Meitner processes or an
$R^{-3}$ dependence for ICD processes within the asymptotic approximation
\cite{Averbukh04,Gokhberg10_1,Fasshauer13,Fasshauer14_1}.
This effect will be
discussed separately in a follow-up article.

Let us put this expression in Eq.~(\ref{eq:z_nucl})
into perspective by comparing it to its
purely electronic counterpart for the decay of one electronic resonance state
to several electronic final states:

\begin{align}
 z_\text{el}(\underline{E}) &=
 \frac{\underline{E} - E_r - \sum\limits_{f}F_{f}(\underline{E})}
 {\sum\limits_f |V_{rf}|^2} \, .
\end{align}
The similarities between them are obvious. The only difference is
weighting the
different decay widths by the Franck-Condon factors
$| \braket{\chi_{\lambda}|\chi_{\mu'}} |^2$.

By again using $\delta_{\lambda,\lambda'}$ consistently in the
normalization of the wavefunction
by calculating $\braket{\Psi_{\underline{E}} | \Psi_E}$
while, analogously to Fano, using
\begin{align}
&\frac{1}{ (\underline{E} - E' - E_{\mu'})(E - E' - E_{\mu'}) }  \nonumber \\
 &= \frac{1}{\underline{E} - E}
   \left( \frac{1}{E-E'-E_{\mu'}} - \frac{1}{\underline{E}-E'-E_{\mu'}} \right)
   \nonumber  \\
 & \quad + \pi^2 \delta(\underline{E} - E) \delta(E' -(E - E_{\mu'}))
\end{align}
\begin{align}
& \delta(\underline{E}-E'-E_{\mu'}) \delta(E-E'-E_{\mu'})  \nonumber \\
 &= \delta(\underline{E} - E) \delta(E' - \frac12 (\underline{E} + E -2E_{\mu'}))
\end{align}

and by assuming that $F(\underline{E})$ is real, we obtain

\begin{equation} \label{eq:norm_a_sum}
	\sum\limits_\lambda |a_\lambda(\underline{E})|^2 W_\lambda (\pi^2 + |z(\underline{E})|^2) \delta(E - \underline{E})
	= \delta(E - \underline{E}) \, .
\end{equation}

To solve this equation for the coefficient of only one vibrational eigenstate of the
resonance state $\lambda$, we assume that the contributions of the different
vibrational eigenstates are equally distributed amongst all $N_\lambda$
vibrational bound states of the resonance state to give

\begin{align} \label{eq:norm_a}
  |a_\lambda(\underline{E})|^2 W_\lambda (\pi^2 + |z(\underline{E})|^2)
 = 1 / N_\lambda \, .
\end{align}

We finally determine the coefficients within the applied approximations to be

%The coefficients, determined by normalization of the ansatz, are given by
\begin{widetext}
\begin{align} \label{eq:Fano_coeff_a}
 a_\lambda(\underline{E})
            =& - \frac{\sqrt{W_\lambda}}
                 {\sqrt{N_\lambda} 
                  \sqrt{(\underline{E}-E_r- E_\lambda
                    - \sum\limits_{\mu''}F_{\mu''}(\underline{E}))^2
                    + \pi^2 W_\lambda^2} } \\
 b_{E',\mu'}(\underline{E})
 =& \sum\limits_{\lambda'}
 \frac{V_{E'r} a_{\lambda'}(\underline{E}) \braket{\chi_{\mu'} | \chi_{\lambda'}}}
 {\underline{E} - E' - E_{\mu'}}
 + \sum\limits_{\lambda'}
 \frac{( \underline{E}-E_r - E_{\lambda'}
 	- \sum_{\mu''}F_{\mu''}(\underline{E})) \quad
 	V_{E'r} a_{\lambda'}}
 {W_{\lambda'}}
 \delta(\underline{E} - E' - E_{\mu'}) 
 \braket{\chi_{\mu'} | \chi_{\lambda'}} . \label{eq:Fano_coeff_b}
\end{align}
\end{widetext}

%\begin{widetext}
%\begin{align} \label{eq:Fano_coeff_a}
% a_\lambda(\underline{E},\langle R \rangle)
%            =& - \frac{\sqrt{W_\lambda}}
%                 {\sqrt{N_\lambda} 
%                  \sqrt{(\underline{E}-E_r- E_\lambda
%                    - \sum\limits_{\mu''}F_{\mu''}(\underline{E}))^2
%                    + \pi^2 W_\lambda^2} } \\
% b_{E',\mu'}(\underline{E}, \langle R \rangle)
% =& \sum\limits_{\lambda'}
% \frac{V_{E'r} a_{\lambda'}(\underline{E}) \braket{\chi_{\mu'} | \chi_{\lambda'}}}
% {\underline{E} - E' - E_{\mu'}}
% + \sum\limits_{\lambda'}
% \frac{( \underline{E}-E_r - E_{\lambda'}
% 	- \sum_{\mu''}F_{\mu''}(\underline{E})) \quad
% 	V_{E'r} a_{\lambda'}}
% {W_{\lambda'}}
% \delta(\underline{E} - E' - E_{\mu'}) 
% \braket{\chi_{\mu'} | \chi_{\lambda'}} , \label{eq:Fano_coeff_b}
%\end{align}
%\end{widetext}
%where the inclusion of the argument $\langle R \rangle$ serves as a reminder
%that the coefficients could in principle also depend on an expectation value of the geometry
%and that we have assumed such an explicit dependence to be nonexistent.

Because the $F_{\mu'}(\underline{E})$ are small shifts in the energy position of the
resonance state $\ket{r}\ket{\chi_\lambda}$ caused by the interaction with the continuum,
which are further damped by Franck-Condon factors, we neglect them in the following.
%%%%%%%%%%%%%%%%%%%%%%%%%%%%%%%%%%%%%%%%%%%%%%%%%%%%%%%%%%%%%%%%%%%%%%%%%%%%%%%%%

\subsection{Derivation of the time-dependent formulation for the description of
            time-resolved electronic decay spectra including nuclear degrees of
            freedom}

The relevant property for the description of the time evolution of ICD processes
is the time- and energy-differential ionization probability obtained from the time-dependent wavefunction $ \ket{\Psi(t)}$ by
\begin{equation} \label{eq:td_probability}
P(E_\text{kin},t) = \sum\limits_i  \big|\braket{E_i|\Psi(t) }\big|^2  \, .
\end{equation}
Here $\ket{E_i}$ denotes a continuum state with energy $E_i$, which entails both the
kinetic energy $E_\text{kin}$ of the emitted electron and the energy of the
final cationic state $i$.
The electronic continuum states
are orthogonal to all electronic bound states of the system and the
nuclear eigenstates of each electronic state are orthogonal amongst
each other.
Atomic units are used throughout unless stated otherwise.

We obtain the wavefunction of the system by
solving the
time-dependent Schrödinger equation
\begin{equation}
 i \frac{\partial}{\partial t} \ket{\Psi(t)} = H(t) \ket{\Psi(t)}
\end{equation}
for the Hamiltonian
\begin{equation}
 H(t) = H_{0} + V  + H_{X}(t)   \, .
\end{equation}

Here, $H_0$ is the Hamiltonian of the unperturbed system, $V$ is the configuration
interaction operator coupling resonance and continuum states and $H_X$ describes
the electro-magnetic field of the exciting laser pulse.

We are going to use the formalism of time-evolution operators $ U(t,t_0)$, which
adhere to a specific Hamiltonian and describe the propagation of a system from time
$t_0$ to time $t$:
\begin{equation}
 \ket{\Psi(t)} = U(t,t_0) \ket{\Psi(t_0)} \, .
\end{equation}
%and that also fulfills the time-dependent Schrödinger equation
%\begin{equation}
% i \frac{\partial}{\partial t} U(t,t_0) = H(t) U(t,t_0) .
%\end{equation}

By using time-dependent perturbation theory, we can split off a small perturbation of
the Hamiltonian from the rest
\begin{align}
U(t,t_0)
            = \tilde{U}(t,t_0) - i \int\limits_{t_0}^t \mathrm{d}t' \,
               \tilde{U}(t,t') \, H^1(t') \, U(t',t_0)  \, ,
 \label{eq:US_form_sol}
\end{align}
where $\tilde{U}(t,t_0)$ is the time-evolution operator of the system described by the
zeroth order Hamiltonian $\tilde{H}$ and $H^1(t')$ is the
first-order perturbation. As we will use this relationship twice and
will redefine what is considered to be part of the unperturbed
Hamilton operator, we start with this general expression.

Since the time-evolution operator
$U(t,t_0)$ appears on both sides of this integral equation, it can be
solved iteratively by repeated insertion of Eq.~(\ref{eq:US_form_sol}) into itself:
\begin{multline} \label{eq:US_first_order}
 U(t,t_0) = \tilde{U}(t,t_0)    \\
 - i \int\limits_{t_0}^{t} \mathrm{d}t' \,
 \tilde{U}(t,t') \,  H^1(t') \tilde{U}(t',t_0)
 + \mathcal{O}\bigl[ (H^1)^2\bigr] .
\end{multline}

The term up to first order is then given by:
\begin{align}\label{eq:wavefct_first_order}
\begin{split}
 \ket{\Psi(t)} = \,
 & \tilde{U}(t,t_0) \ket{\Psi(t_0)}   \\
 &- i \int\limits_{t_0}^t \mathrm{d}t' \, \tilde{U}(t,t') \, H^1(t')
 \tilde{U}(t',t_0) \ket{\Psi(t_0)} \, .
\end{split}
\end{align}

%--------------------------------------------------------------

We assume that the intensity of the exciting laser is so low that its
interaction with the system can be described within the dipole approximation.
Therefore, we can describe our system within first order of perturbation theory 
in $H_X(t')$. Starting from the ground state $\ket{G(t_0)}$,
our wavefunction is given by
\begin{align}\label{eq:first_order_psi}
\begin{split}
 \ket{\Psi(t)} = \,
 & \tilde{U}(t,t_0) \ket{G(t_0)}  \\
 &-i \int\limits_{t_0}^{t} \mathrm{d}t' \, \tilde{U}(t,t') H_X(t')
 \tilde{U}(t',t_0) \ket{G(t_0)} \, .
\end{split}
\end{align}
We are free to choose an energy reference and therefore select $E_G=0$.
This conveniently
removes the time dependence from the ground state, since
$\ket{G(t_0)} = \exp\{ -i E_G t_0 \} \ket{G} = \ket{G}$.

We introduce approximations for the time-evolution operator $\tilde{U}(t,t')$
appearing in the last term of Eq.~(\ref{eq:first_order_psi}).
After the system has interacted with the XUV via $H_X(t')$ at time $t'$,
the system decays via the configuration interaction $V$.
In the continuum, we neglect the Coulomb interaction for simplicity,
which leads to  the time evolution operator 
\begin{equation} \label{eq:Utilde}
 \tilde{U}(t, t') = U_0(t,t') - i \int\limits_{t'}^t \mathrm{d}t'' \,
                                  U_0(t,t'') \, V \, \tilde{U}(t'',t')  \, ,
\end{equation}
where  $U_0(t,t')$ is the free-particle time-evolution
operator.

By inserting Eq.~(\ref{eq:Utilde}) into Eq.~(\ref{eq:first_order_psi}) we arrive at
\begin{align}
\begin{split}
 \ket{\Psi(t)} = \,
                & \tilde{U}(t,t_0) \ket{G} \\
                & -i \int\limits_{t_0}^{t} \mathrm{d}t' \,
                  U_0(t,t') H_X(t') \ket{G}\\
                & -\int\limits_{t_0}^{t} \mathrm{d}t'
                   \int\limits_{t'}^t \mathrm{d}t'' \,
                  U_0(t,t'') \, V \, \tilde{U}(t'',t') \, H_X(t') \ket{G} .
\end{split}
\end{align}

The time-evolution operator $\tilde{U}(t'',t')$ in the last integral describes the
contribution of the Hamiltonian $H_0 + V$.
This is analogous to the Hamilton
operator used in Fano's description \cite{Fano61} of a decaying resonance state and equivalent to the operator
in Eq.~(\ref{eq:Fano_Hamiltonian}).
We therefore add the
subscript $\mathrm{F}$ to the time-evolution operator and remove the tilde.

We now introduce the separation of electronic and nuclear wavefunctions. For the ground
state, we assume the following product ansatz of the electronic wavefunction
$\ket{g}$ and the vibrational-ground-state wavefunction $\ket{\chi_0}$ :
\begin{equation}
 \ket{G} = \ket{g} \ket{\chi_0} .
\end{equation}

In order to determine the probability amplitude for one specific final state,
we project on one electronic continuum state $\ket{E}$ and one vibrational state
$\ket{\chi_\mu}$ of this final state, characterized by their respective energies
of the ionized system and the kinetic energy of the emitted electron.
Inserting the resolution of the identity behind the time-evolution operator $U_0$
in both the second and the third term on the right-hand side yields:

\begin{widetext}
\begin{align}
\begin{split}
% \bra{\chi_\mu}\braket{E |\Psi(t)} =& \bra{\chi_\mu} \braket{E | \tilde{U}(t,t_0)| g }\ket{\chi_0} \\
%                & -i \int\limits_{t_0}^{t} \mathrm{d}t' \,
%                  \bra{\chi_\mu} \braket{E |U_0(t,t') H_X(t')| g} \ket{\chi_0}\\
%                & -\int\limits_{t_0}^{t} \mathrm{d}t'
%                    \int\limits_{t'}^t \mathrm{d}t'' \,
%                  \bra{\chi_\mu} \braket{E| U_0(t,t'') \, V \, \tilde{U}(t'',t') \, H_X(t') |g} \ket{\chi_0} .
 \bra{\chi_\mu}\braket{E |\Psi(t)} =
 &\bra{\chi_\mu} \braket{E | U_\mathrm{F}(t,t_0)| g }\ket{\chi_0} \\
 &                 -i \int\limits_{t_0}^{t} \mathrm{d}t' \,
                  \bra{\chi_\mu} \braket{E |U_0(t,t')
                   H_X(t')| g} \ket{\chi_0}\\
 &                 -\int\limits_{t_0}^{t} \mathrm{d}t'
                    \int\limits_{t'}^t \mathrm{d}t'' \,
                  \bra{\chi_\mu} \braket{E| U_0(t,t'') \, V \, U_\mathrm{F} (t'',t') \, H_X(t') |g} \ket{\chi_0} \\
 =&\bra{\chi_\mu} \braket{E | U_\mathrm{F}(t,t_0)| g }\ket{\chi_0} \\
 &                 -i \int\limits_{t_0}^{t} \mathrm{d}t'
                      \sum_{\mu'} \int \mathrm{d}E' \,
                  \bra{\chi_\mu} \braket{E |U_0(t,t') | E'} \ket{\chi_{\mu'}}
                  \bra{\chi_{\mu'}}\braket{E' | H_X(t')| g} \ket{\chi_0}\\
 &                 -\int\limits_{t_0}^{t} \mathrm{d}t'
                    \int\limits_{t'}^t \mathrm{d}t'' 
                      \sum_{\mu'} \int \mathrm{d}E' \,
                  \bra{\chi_\mu} \braket{E| U_0(t,t'') | E'} \ket{\chi_{\mu'}}
                  \bra{\chi_{\mu'}} \braket{E' | V \, U_\mathrm{F} (t'',t') \, H_X(t') |g} \ket{\chi_0} \\
\label{eq:first_projection}
\end{split}
\end{align}
\end{widetext}

The time-evolution operator $U_0(t,t')$  solves the time-dependent
Schrödinger equation
for the Hamiltonian $\vec{p}^2/2 + E_\text{fin} + E_\mu$.
The corresponding wavefunctions with
wavevector $\vec{k}$ are
$\ket{\Psi_{\vec{k}}^0 (t)}
  = \ket{\vec{k}}
    \exp\biggl[ -i \int\limits_{-\infty}^t \mathrm{d}t''' (\frac 12 \vec{k}^2  + E_\text{fin} + E_\mu)\biggr]$.
Its projection formulation propagating a system from time $t'$ to time $t$ is given by

\small
\begin{align}
\begin{split}
 U_0(t,t') &= 
           \int \mathrm{d}\vec{k} \, \ket{\Psi_{\vec{k}}^0 (t)} \bra{\Psi_{\vec{k}}^0 (t')} \\
%           &= U_\text{e}(t,t') \otimes U_\text{nucl}(t,t') \\
           &= \sum_{\tilde{\mu}} \int \mathrm{d} \tilde{E}
              \ket{\tilde{E}(t)} \ket{\chi_{\tilde{\mu}}(t)}
              \bra{\chi_{\tilde{\mu}}(t')} \bra{\tilde{E}(t')} \\
           &= \sum\limits_{\tilde{\mu}} \int \mathrm{d} \tilde{E}
              \exp \biggl[ -i \int\limits_{t'}^t \mathrm{d} t''' (\tilde{E} + E_{\tilde{\mu}})
             \biggr]
              \ket{\tilde{E}} \ket{\chi_{\tilde{\mu}}}
              \bra{\chi_{\tilde{\mu}}} \bra{\tilde{E}} \, .
\end{split}
\end{align}
\normalsize

At this point we only consider the kinetic energy of the emitted electron
and can therefore
write the projection of a general continuum state $\bra{E}$ on the free-particle
time-evolution operator in the following way:

\small
\begin{align}
 &\bra{\chi_\mu} \braket{E| U_0(t,t') |E'} \ket{\chi_{\mu'}} \nonumber \\
						  &=\int \mathrm{d}\tilde{E}
						  	 \sum_{\tilde{\mu}}
						  	 \bra{\chi_\mu} \braket{E| \, \exp[ i\Phi_0 (E,t,t') ]
						  	 | \tilde{E}} \ket{\chi_{\tilde{\mu}}} \nonumber \\
						  	 & \quad \times \bra{\chi_{\tilde{\mu}}} \bra{\tilde{E}}  
                            \ket{E'} \ket{\chi_{\mu'}}\\
                          &=\int \mathrm{d}\tilde{E}
                             \sum_{\tilde{\mu}}
                             \bra{\chi_\mu} \bra{E}
                             \ket{\tilde{E}} \ket{\chi_{\tilde{\mu}}}
                             \, \exp \biggl[ -i \int\limits_{t'}^{t} \mathrm{d}t'''(\tilde{E} + E_{\tilde{\mu}})  \biggr]
                             \nonumber\\
                          &  \quad  \times \bra{\chi_{\tilde{\mu}}} \bra{\tilde{E}}
                             \ket{E'} \ket{\chi_{\mu'}} \nonumber\\
                          %=& 
                          %   \sum_{\tilde{\mu}}
                          %   \, \exp \biggl[ -i \int\limits_{t'}^{t}
                          %   E_\text{kin} + E_\text{fin}
                          %   + E_{\tilde{\mu}} \, \mathrm{d}t''' \biggr]
                          %   \delta(E-E') \delta_{\mu,\mu'}
                          %   \\
                          &= 
                          	 \exp \biggl[ -i \int\limits_{t'}^{t} \mathrm{d}t'''
                           (E_\text{kin} + E_\text{fin}
                          	 + E_{\mu})  \biggr]
                         \, \delta(E-E') \, \delta_{\mu,\mu'}    \;,
\end{align}
\normalsize
where $E_\text{kin}$ is the kinetic energy relative to the final cationic state
$E_\text{fin} + E_\mu$ and $\Phi_0 (E,t,t')$ is the time-dependent phase.

The first term of the right-hand side of Eq.~(\ref{eq:first_projection}) vanishes
because of the orthogonality between electronic bound and continuum eigenstates.
The non-zero
terms of Eq.~(\ref{eq:first_projection}) are given by

\begin{widetext}
\begin{align}
\begin{split}
	\bra{\chi_\mu} \braket{E| \Psi(t)} 
 %    = & -i \sum\limits_{\tilde{\mu}} \int \mathrm{d}\tilde{E}
	% \int\limits_{t_0}^{t} \mathrm{d}t' \,
	% \bra{\chi_\mu} \braket{E|\tilde{E}} \ket{\chi_{\tilde{\mu}}}
	% \exp \biggl[ -i (t-t') (\tilde{E} + E_{\tilde{\mu}}) \biggr]
	% \bra{\chi_{\tilde{\mu}}} \braket{\tilde{E} |H_X(t')| g} \ket{\chi_0}\\
	% & - \sum\limits_{\tilde{\mu}} \int \mathrm{d}\tilde{E}
	% \int\limits_{t_0}^{t} \mathrm{d}t'
	% \int\limits_{t'}^t \mathrm{d}t'' \,
	% \bra{\chi_\mu} \braket{E|\tilde{E}} \ket{\chi_{\tilde{\mu}}}
	% \exp \biggl[ -i (t-t'') (\tilde{E} + E_{\tilde{\mu}}) \biggr]
	% \bra{\chi_{\tilde{\mu}}} \braket{\tilde{E}|V \, U_\mathrm{F} (t'',t') \, H_X(t') |g}\ket{\chi_0} \\
              = & -i 
                     \int\limits_{t_0}^{t} \mathrm{d}t' \,
                     \exp \biggl[ -i (t-t') (E_\text{kin} + E_\text{fin} + E_{\mu}) \biggr]
                   \bra{\chi_{\mu}} \braket{E |H_X(t')| g} \ket{\chi_0}\\
                & - 
                   \int\limits_{t_0}^{t} \mathrm{d}t'
                   \int\limits_{t'}^t \mathrm{d}t'' \,
                     \exp \biggl[ -i (t-t'') (E_\text{kin} + E_\text{fin} + E_{\mu}) \biggr]
                  \bra{\chi_{\mu}} \braket{E|V \, U_\mathrm{F} (t'',t') \, H_X(t') |g}\ket{\chi_0}   \, .
\end{split}
\end{align}
\end{widetext}

The second term can be solved by substituting the configuration interaction operator
\begin{align}\begin{split}
\label{eq:V}
 \hat{V} =& \sum_{\mu'} \sum_{\lambda} \int \mathrm{d}E' \,
           \ket{E'} \ket{\chi_{\mu'}} \braket{\chi_{\mu'}|\chi_\lambda}
           V_{E'r} \bra{\chi_\lambda} \bra{r}\\
         &+ \sum_{\mu'} \sum_{\lambda} \int \mathrm{d}E' \, 
           \ket{r} \ket{\chi_\lambda} \braket{\chi_\lambda | \chi_{\mu'}}
           V_{rE'} \bra{\chi_{\mu'}} \bra{E'} 
\end{split}\end{align}
and inserting the resolution of the identity
\begin{align}\begin{split}
 \mathbb{1} =& \sum_{\lambda'} \ket{r} \ket{\chi_{\lambda'}} \bra{\chi_{\lambda'}} \bra{r}\\
             &+ \sum_{\mu''} \int \mathrm{d}E'' \,
              \ket{E''}\ket{\chi_{\mu''}} \bra{\chi_{\mu''}} \bra{E''}
\end{split}\end{align}
between the Fano propagator and the operator of the exciting field to yield

\begin{widetext}
\small
\begin{align}
\label{eq:before_Fano}
\begin{split}
 & \bra{\chi_\mu} \braket{E| \Psi(t)} =\\
                & -i 
                     \int\limits_{t_0}^{t} \mathrm{d}t' \,
                     \exp \biggl[ -i (t-t') (E_\text{kin} + E_\text{fin} + E_{\mu}) \biggr]
                   \bra{\chi_{\mu}} \braket{E |H_X(t')| g} \ket{\chi_0}\\
                & -\sum\limits_{\lambda,\lambda'} 
                   \int\limits_{t_0}^{t} \mathrm{d}t'
                   \int\limits_{t'}^t \mathrm{d}t'' \,
                     \exp \biggl[ -i (t-t'') (E_\text{kin} + E_\text{fin} + E_{\mu}) \biggr]  
                      V_{Er} \braket{\chi_\mu | \chi_\lambda}
                  \bra{\chi_\lambda} \braket{r| U_\mathrm{F} (t'',t') | r} \ket{\chi_{\lambda'}}
                  \, \bra{\chi_{\lambda'}} \braket{r |H_X(t') |g}\ket{\chi_0} \\
                & -\sum\limits_{\lambda,\mu''} \int \mathrm{d} E''
                   \int\limits_{t_0}^{t} \mathrm{d}t'
                   \int\limits_{t'}^t \mathrm{d}t'' \,
                     \exp \biggl[ -i (t-t'') (E_\text{kin} + E_\text{fin} + E_{\mu}) \biggr] 
                     V_{Er} \braket{\chi_\mu | \chi_\lambda}
                  \bra{\chi_\lambda} \braket{r| U_\mathrm{F} (t'',t') | E''} \ket{\chi_{\mu''}}
                  \, \bra{\chi_{\mu''}} \braket{E'' |H_X(t') |g}\ket{\chi_0}    \, .
\end{split}
\end{align}
\end{widetext}

% Here, we have assumed that the transition dipole matrix element from the
% ground state to the continuum
% are independent of the energy of the continuum state,
% $\braket{E_i|\mu|g} = \braket{c|\mu|g}$,
% over the energy range of interest.

Since we describe the interaction between the system and the exciting XUV field
in the dipole approximation,
the corresponding  Hamilton operator in the length gauge is given by
$H_X(t') = -\mu \mathcal{E}_X(t') = \mu \frac{\text{d}}{\text{d}t'} A_X(t') f_X(t')$,
where $\mathcal{E}_X(t')$ denotes the time-dependent field strength of the XUV laser,
while $\mu$ denotes the dipole operator, $A_X(t') = A_{0X} \cos(\Omega t')$
is the vector potential of the laser
field in the direction of the linear polarization
and $f_X(t')$ is the Gaussian pulse envelope.
We assume the transition dipole matrix element from the ground state to the continuum
to be independent of the energy of the continuum state,
\begin{equation}
\braket{E_i|\mu|g} = \braket{c|\mu|g},
\end{equation}
over the energy range of interest:

\begin{widetext}
\small
\begin{align}
\label{eq:before_Fano_dipole}
\begin{split}
 & \bra{\chi_\mu} \braket{E| \Psi(t)} =\\
                & +i \braket{c|\mu|g}
                     \int\limits_{t_0}^{t} \mathrm{d}t' \,
                     \exp \biggl[ -i (t-t') (E_\text{kin} + E_\text{fin} + E_{\mu}) \biggr]
                   \braket{\chi_{\mu} | \chi_0} \mathcal{E}_X(t')\\
                & + \braket{r|\mu|g} \sum\limits_{\lambda,\lambda'} 
                   \int\limits_{t_0}^{t} \mathrm{d}t'
                   \int\limits_{t'}^t \mathrm{d}t'' \,
                     \exp \biggl[ -i (t-t'') (E_\text{kin} + E_\text{fin} + E_{\mu}) \biggr]  
                      V_{Er} \braket{\chi_\mu | \chi_\lambda}
                  \bra{\chi_\lambda} \braket{r| U_\mathrm{F} (t'',t') | r} \ket{\chi_{\lambda'}}
                  \braket{\chi_{\lambda'} | \chi_0} \mathcal{E}_X(t') \\
                & + \braket{c|\mu|g} \sum\limits_{\lambda,\mu''} \int \mathrm{d} E''
                   \int\limits_{t_0}^{t} \mathrm{d}t'
                   \int\limits_{t'}^t \mathrm{d}t'' \,
                     \exp \biggl[ -i (t-t'') (E_\text{kin} + E_\text{fin} + E_{\mu}) \biggr] 
                     V_{Er} \braket{\chi_\mu | \chi_\lambda}
                  \bra{\chi_\lambda} \braket{r| U_\mathrm{F} (t'',t') | E''} \ket{\chi_{\mu''}}
                  \braket{\chi_{\mu''} | \chi_0} \mathcal{E}_X(t')    \, .
\end{split}
\end{align}
\end{widetext}

The Fano matrix elements including nuclear dynamics require a thorough inspection.
We assume a Born-Oppenheimer ansatz of the wavefunction,
in which the electronic solutions
are parametrically dependent on the nuclear coordinates and the coefficients do not
directly depend on either the electronic or nuclear coordinates.
It is therefore
possible to use the following projection formulation:

\small
\begin{widetext}
\begin{align}
 U_\mathrm{F}(t'',t') =& \int \mathrm{d}\underline{E} \, \ket{\Psi_{\underline{E}}(t'')}
                        \bra{\Psi_{\underline{E}}(t')} \nonumber\\
             =& \int \mathrm{d}\underline{E} \,
                \ket{\Psi_{\underline{E}}}
                \bra{\Psi_{\underline{E}}} \,
                \exp[-i\underline{E} (t''-t')] \nonumber\\
             =& \int \mathrm{d}\underline{E} \,
                \exp[-i\underline{E} (t''-t')]
                \biggl[
                  \sum\limits_{\lambda'',\lambda'''}
                  a_{\lambda''}(\underline{E}) a^*_{\lambda'''}(\underline{E})
                  \ket{r} \ket{\chi_{\lambda''}}  \bra{\chi_{\lambda'''}} \bra{r}
                 + \sum\limits_{\lambda'',\tilde{\mu}} \int \mathrm{d}\tilde{E} \,
                  a_{\lambda''}(\underline{E}) b^*_{\tilde{E},\tilde{\mu}}(\underline{E})
                  \ket{r} \ket{\chi_{\lambda''}}  \bra{\chi_{\tilde{\mu}}} \bra{\tilde{E}}
                 + \cdots
                \biggr] \nonumber\\
             =& \int \mathrm{d}\underline{E} \,
                \exp[-i\underline{E} (t''-t')]
                \biggl[
                  \sum\limits_{\lambda''}
                  a_{\lambda''}(\underline{E}) a^*_{\lambda''}(\underline{E})
                  \ket{r} \ket{\chi_{\lambda''}}  \bra{\chi_{\lambda''}} \bra{r}
                 + \sum\limits_{\lambda'',\tilde{\mu}} \int \mathrm{d}\tilde{E} \,
                  a_{\lambda''}(\underline{E}) b^*_{\tilde{E},\tilde{\mu}}(\underline{E})
                  \ket{r} \ket{\chi_{\lambda''}}  \bra{\chi_{\tilde{\mu}}} \bra{\tilde{E}}
                 + \cdots
                \biggr]
 \, .   \label{eq:Fano_projection}
\end{align}
\end{widetext}
\normalsize
Because we assume that there is no coupling between nuclear eigenstates
of the same electronic
eigenstate via the continuum,
$\lambda''$ needs to be equal
to $\lambda'''$.
% The same logic applies when evaluating
%$b^*_{\tilde{E},\tilde{\mu}}(\underline{E})$ in the derivation of the coefficients
%in Appendix \ref{app:Fano_coeff}, in which a Franck-Condon overlap integral
%is included.

As already mentioned, the derivation of the coefficients of
Eqs.~(\ref{eq:Fano_coeff_a}, \ref{eq:Fano_coeff_b}) 
%The determination of the coefficients given in Appendix \ref{app:Fano_coeff}
relies on
several, in part crude, approximations. For a quantitative comparison to
experiment, these should
be checked carefully. However, they are feasible for this qualitative description
of the influence of nuclear dynamics on time-resolved spectroscopy on electronic
decay processes.
The Fano integrals can then be solved 
by contour integration in the negative complex half-plane as shown in 
Appendix~\ref{app:Fano_me}. Inserting them into Eq.~(\ref{eq:before_Fano_dipole})
yields our working equation:
% 
% 
% \begin{widetext}
% \begin{align}
% \begin{split}\label{eq:ampl_after_Fano}
%  \bra{\chi_\mu} \braket{E| \Psi(t)} =
%                 & -i 
%                      \int\limits_{t_0}^{t} \mathrm{d}t' \,
%                      \exp \biggl[ -i (t-t') (E_\text{kin} + E_\text{fin} + E_{\mu}) \biggr]
%                    \bra{\chi_{\mu}} \braket{E |H_X(t')| g} \ket{\chi_0}\\
%                 & - \frac{1}{N_\lambda}  \sum\limits_{\lambda} 
%                    \int\limits_{t_0}^{t} \mathrm{d}t'
%                    \int\limits_{t'}^t \mathrm{d}t'' \,
%                      \exp \biggl[ -i (t-t'') (E_\text{kin} + E_\text{fin} + E_{\mu}) \biggr]    \\
%                 & \qquad  \times V_{r} \braket{\chi_\mu | \chi_\lambda}
%                   \braket{\chi_\lambda | \chi_0}
%                      \exp \biggl[ -i (t''-t') (E_r + E_\lambda - i\pi W_\lambda) \biggr] 
%                   \, \braket{r |H_X(t') |g} \\
%                 & + \frac{i\pi}{N_\lambda}
%                    \sum\limits_{\lambda,\mu''}
%                    \int\limits_{t_0}^{t} \mathrm{d}t'
%                    \int\limits_{t'}^t \mathrm{d}t'' \,
%                      \exp \biggl[ -i (t-t'') (E_\text{kin} + E_\text{fin} + E_{\mu}) \biggr]    \\
%                 & \qquad  \times |V_{r}|^2 \braket{\chi_\mu | \chi_\lambda}
%                   \braket{\chi_\lambda | \chi_{\mu''}} \braket{\chi_{\mu''} | \chi_0}
%                      \exp \biggl[ -i (t''-t') (E_r + E_\lambda - i\pi W_\lambda) \biggr] 
%                   \, \braket{E |H_X(t') |g} \, .
% \end{split}
% \end{align}
% \end{widetext}
% 
% 
% 
% 
% 
% Hence, our working equation is:

\begin{widetext}
\begin{align}
\label{eq:ampl_working_eq}
\begin{split}
 \bra{\chi_\mu} \braket{E| \Psi(t)} =
                & +i \braket{c|\mu|g}
                     \int\limits_{t_0}^{t} \mathrm{d}t' \,
                     \exp \biggl[ -i (t-t') (E_\text{kin} + E_\text{fin} + E_{\mu}) \biggr] \,
                    \braket{\chi_{\mu}| \chi_0} \mathcal{E}_X(t')\\
                & + \frac{\braket{r|\mu|g}}{N_\lambda}  \sum\limits_{\lambda} 
                   \int\limits_{t_0}^{t} \mathrm{d}t'
                   \int\limits_{t'}^t \mathrm{d}t'' \,
                     \exp \biggl[ -i (t-t'') (E_\text{kin} + E_\text{fin} + E_{\mu}) \biggr]    \\
                & \qquad  \times V_{r} \braket{\chi_\mu | \chi_\lambda}
                  \braket{\chi_\lambda | \chi_0}
                     \exp \biggl[ -i (t''-t') (E_r + E_\lambda - i\pi W_\lambda) \biggr] 
                  \, \mathcal{E}_X(t') \\
                & - \frac{i\pi \braket{c|\mu|g}}{N_\lambda}
                   \sum\limits_{\lambda,\mu''}
                   \int\limits_{t_0}^{t} \mathrm{d}t'
                   \int\limits_{t'}^t \mathrm{d}t'' \,
                     \exp \biggl[ -i (t-t'') (E_\text{kin} + E_\text{fin} + E_{\mu}) \biggr]    \\
                & \qquad  \times |V_{r}|^2 \braket{\chi_\mu | \chi_\lambda}
                  \braket{\chi_\lambda | \chi_{\mu''}} \braket{\chi_{\mu''} | \chi_0}
                     \exp \biggl[ -i (t''-t') (E_r + E_\lambda - i\pi W_\lambda) \biggr] 
                  \, \mathcal{E}_X(t')  \, .\\
\end{split}
\end{align}
\end{widetext}

%
%\begin{align}
%\begin{split}
% \bra{\chi_\mu} \braket{E| \Psi(t)} =
%                & -i \braket{\chi_\mu | \chi_0} \braket{c|\mu|g}
%                     \int\limits_{t_0}^{t} \mathrm{d}t' \,
%                     \mathcal{E}_X(t')
%                     \exp \biggl[ -i (t-t') (E_\text{kin} + E_\text{fin} + E_{\mu}) \biggr] \\
%                & + \biggl[ \frac{V_r \braket{r| \mu |g}}{N_\lambda}
%                   - \frac{i\pi W V_r \braket{c |\mu| g}}{N_\lambda} \biggr]
%                  \sum\limits_{\lambda} 
%                  \braket{\chi_\mu | \chi_\lambda} \braket{\chi_\lambda | \chi_0} \\
%                & \quad  \times 
%                   \int\limits_{t_0}^{t} \mathrm{d}t'
%                   \int\limits_{t'}^t \mathrm{d}t'' \,
%                     \mathcal{E}_X(t')
%                     \exp \biggl[ -i (t-t'') (E_\text{kin} + E_\text{fin} + E_{\mu}) \biggr]  
%                     \exp \biggl[ -i (t''-t') (E_r + E_\lambda - i\pi W) \biggr] 
%\end{split}
%\end{align}
%

%Here, we assumed that the exciting pulse is not able to initiate the ICD process.
%Eq.~(\ref{eq:ampl_three_S}),  Eq.~(\ref{eq:ampl_three}) in the main text,
%contains one more term than Eq.~(11) from Ref.~\cite{Wickenhauser05} due to the
%second possible final state.
The first part describes the direct
ionization into the continuum final state,
while the second part of Eq.~(\ref{eq:ampl_working_eq})
describes the excitation from the ground state to the resonance state
followed by a decay
to the continuum final state.
The third part describes the excitation to
a continuum state, which then couples to the resonance state, which then again decays
into either the same or a different continuum final state.

The first integral thus corresponds to the direct ionization process. The second (double) integral
is identical for the resonance and the indirect terms. They have, though, different
prefactors and are weighted differently for the nuclear states.
For the case of a slowly varying envelope function
($\frac{\text{d}}{\text{d}t'} f_X(t') \approx 0$) and only considering the
absorption of an XUV photon (neglecting the emission), these two integrals are solved analytically for times
after the exciting pulse.
The direct term for a specific value of $\mu$ is given by
\small
\begin{widetext}
\begin{align}
\begin{split}
 - \frac{A_{0X} \Omega \braket{c|\mu|g}}{2} \braket{\chi_\mu|\chi_0} \,
   \exp\biggl[ -it(E_{\text{kin}} + E_{\text{fin}} + E_\mu) \biggr] \,
   \exp\biggl[ -\frac{\sigma^2}{2} (E_{\text{kin}} + E_{\text{fin}} + E_{\mu} -\Omega)^2 \biggr] \, \\
   \times \Re \Biggl[  \erf
   \biggl( \frac{1}{\sqrt{2} \sigma} \Bigl( \frac{T_X}{2}
   + i\sigma^2 \bigl( E_{\text{kin}} + E_{\text{fin}} + E_\mu - \Omega \bigr)
   \Bigr) \biggr) \Biggr] \, ,
   \label{eq:dir_ampl}
\end{split}
\end{align}
\end{widetext}
\normalsize
where $\Re$ denotes the real part and $\erf$ denotes the error function.
The resonance and indirect ionization terms without their respective prefactors and
weights
are given by
\begin{widetext}
\small
\begin{alignat}{2}
 &- \frac{A_{0X} \, \Omega}
         {4 (E_r + E_\lambda - i\pi W_\lambda - E_{\text{kin}} - E_{\text{fin}} - E_\mu)} \,
   &&\exp\biggl[ -it(E_r + E_\lambda- i\pi W_\lambda) \biggr] \, \nonumber \\
   &&& \times \exp\biggl[ -\frac{\sigma^2}{2} (E_r + E_\lambda -i\pi W_\lambda -\Omega)^2 \biggr] \,
   \Biggl( \erf(\tau_\text{1,max}) - \erf(\tau_\text{1,min}) \Biggr)             \nonumber\\
 &+ \frac{A_{0X} \, \Omega}
         {4 (E_r + E_\lambda -i\pi W_\lambda - E_{\text{kin}} - E_{\text{fin}} - E_\mu)} \,
   &&\exp\biggl[ -it(E_{\text{kin}} + E_{\text{fin}} + E_\mu) \biggr]\, \nonumber \\
   &&& \times \exp\biggl[ -\frac{\sigma^2}{2} (E_{\text{kin}} + E_{\text{fin}} + E_\mu -\Omega)^2 \biggr] \,
   \Biggl( \erf(\tau_\text{2,max}) - \erf(\tau_\text{2,min}) \Biggr)
   \label{eq:res_ampl}
\end{alignat}
\end{widetext}
with
$\tau_\text{1,max} = \frac{1}{\sqrt{2}\sigma}
 \Bigl( \frac{T_X}{2} - i\sigma^2(E_r + E_\lambda -i\pi W_\lambda - \Omega) \Bigr)$,
$\tau_\text{1,min} = -\frac{1}{\sqrt{2}\sigma}
 \Bigl( \frac{T_X}{2} + i\sigma^2(E_r + E_\lambda -i\pi W_\lambda - \Omega) \Bigr)$,
$\tau_\text{2,max} = \frac{1}{\sqrt{2}\sigma}
 \Bigl( \frac{T_X}{2} - i\sigma^2(E_{\text{kin}} + E_{\text{fin}} +E_\mu - \Omega) \Bigr)$ and
$\tau_\text{2,min} = -\frac{1}{\sqrt{2}\sigma}
 \Bigl( \frac{T_X}{2} + i\sigma^2(E_{\text{kin}} + E_{\text{fin}} + E_\mu - \Omega) \Bigr)$.
Here, $\sigma$ is the standard deviation of the Gauss distribution in time and $T_X$ is
the duration of the exciting XUV pulse. The different $\tau$'s are the
upper and lower limits
of the integral after the coordinate transformation that allows to formally solve
the outer integral of the resonance and the indirect terms in Eq.~(\ref{eq:ampl_working_eq}).

The absolute square of the sum of all these terms
gives the ionization probability for one
specific nuclear final state. In order to determine the total spectrum, the
contributions of all final states need to be added up according to
Eq.~(\ref{eq:td_probability}).

%%%%%%%%%%%%%%%%%%%%%%%%%%%%%%%%%%%%%%%%%%%%%%%%%%%%%%%%%%%%%%%%%%%%%%%%%%%%%%%%%
%\input{subsecs/computational}

\section{Computational Details}
\label{sec:comp}

In the following, we will show the results for different cases
with different numbers of nuclear bound states of either the resonance state
or the final state, which were all obtained by the script \textsc{nuclear\_dyn} within
a locally modified version of \textsc{ELDEST} \cite{ELDEST_v2.0.3, figshare_input_2024} (hash: 7dd1601).

We use Morse potentials in
the simulations:

\begin{equation} \label{eq:Morse}
 V_\text{Morse} = D \left( 1 - \exp(-\alpha R) \right)^2.
\end{equation}

These are beneficial for our description because
the vibrational wavefunctions are known analytically
\cite{Morse29,Scholz32,Pekeris34}
and the
evaluation of the Franck-Condon integrals is therefore
possible without prior numerical
solution of the nuclear Schrödinger equation.
In our implementation, we use the recursive definition given in Eq.~(49) of
Ref.~\cite{Dahl88}. In order to compensate for the integration over $R$ instead of
$\alpha R$ used in Ref.~\cite{Dahl88}, we included an
additional factor of $\sqrt{\alpha}$ in the normalization constant.

However, the general findings, which we will discuss later, will also hold for
other types of interatomic potentials.

Unless stated otherwise, the parameters of the Morse potentials
were optimized for a given number of
bound states, of which as many as possible have significant
Franck-Condon overlaps between
a) the ground and resonance state and
b) the resonance and final state
for equal equilibrium distances.
Moreover, the energy differences between the vibrational ground state
and first excited state are of the size usually found
in existing chemical systems, especially diatomic molecules.
We used $\alpha_\text{max} = 30.0$ and
$D_{\text{max}} = \qty{6.0}{eV}$ and a purely electronic
Fano $q$ parameter of 1 [see Eq.~(\ref{eq:Fano_param})].
The electronic ground state is chosen such that it only
contains one nuclear bound state
and its parameters are kept constant.
All parameters can be found in Table~\ref{tab:params}.

\begin{table}[h]
\centering
\caption{Parameters of the Morse potentials defined in Eq.~(\ref{eq:Morse})
and used for the simulations.}
\label{tab:params}
\begin{tabular}{rrrrrrr}
\toprule
\# & $\alpha_\text{res} [\text{a.u.}]$ & $D_\text{res} [\text{eV}]$ &
     $\alpha_\text{fin} [\text{a.u.}]$ & $D_\text{fin} [\text{eV}]$ \\
\midrule
 1 &  4.9512 & 0.01 & 21.6296 & 0.11 \\% 1l1m
 2 & 21.6296 & 0.11 & 17.0028 & 0.51 \\% 1l2m
 3 &  4.9512 & 0.01 &  4.9121 & 0.81 \\% 1lmm
 4 & 15.3994 & 0.50 & 21.7512 & 0.10 \\% 2l1m
 5 &  4.4414 & 0.04 & 23.9911 & 0.12 \\% 2l1m small dE
 6 & 10.8894 & 0.50 &  5.1507 & 0.10 \\% 2l2m 1.
 7 &  5.1507 & 0.10 & 10.8894 & 0.50 \\% 2l2m 2.
\bottomrule
\end{tabular}
\end{table}

We choose all configuration interaction terms $V_r$
to be real
and to correspond to a lifetime of
$\tau = \frac 1{2\pi |V_r|^2} = \qty{20}{fs}$
to allow for an easier comparison between the different cases.

The electronic energies of the resonance states are chosen such that
the energy difference between the vibrational ground state of the electronic
ground state and the vibrational ground state of the resonance state has an energy
of \qty{50}{eV}. This allows us to use the same parameters for 
the initiating pulse, which
is chosen to have a mean photon energy of \qty{50}{eV} and $n_X=10$ cycles
within the full width at half maximum (FWHM) of the Gaussian pulse. This results in
an FWHM of \qty{1.2}{fs}. The pulse is modelled for a total of 
$\qty{5}{\sigma}$.
Moreover, the electronic final state is chosen such that
the secondary electron emitted during a transition from
the vibrational ground states of the resonance state
to the vibrational ground state of the final state is \qty{10}{eV}.

In order to extract oscillation frequencies from the calculated time-resolved spectra, we employed a fast Fourier transform (FFT).
To this end, we evaluated the signal for one specific kinetic energy,
discarded the first eight data points to exclude the build-up phase from the signal
and performed the FFT on the remaining data points.
The central frequency of a peak was chosen as the centre of its FWHM.

%%%%%%%%%%%%%%%%%%%%%%%%%%%%%%%%%%%%%%%%%%%%%%%%%%%%%%%%%%%%%%%%%%%%%%%%%%%%%%%%%
%\input{subsecs/cases}

\section{Results}
\label{sec:cases}

The basic findings for the purely electronic case
have already been reported
elsewhere \cite{Fasshauer20_1}. We summarize them here, to allow a better comparison
of the following showcases to the hitherto known case.

The shape of the signal after long times is a
Fano profile convoluted with a Gaussian.
The former's shape is described by the ratio between the contribution of the resonance
part to the direct part of the wavefunction, which are
included in the Fano parameter
\begin{equation}\label{eq:Fano_param}
 q=\frac{\braket{r|\mu|g}}{ \braket{c|\mu|g} \pi V^*_{Er} } \, ,
\end{equation}
so that the Fano profile may be described as
\begin{equation}\label{eq:Fano_shape_el}
 \frac{\left| \braket{\Psi_{\underline{E}} | \mu | g} \right|^2}
 {\left| \braket{\underline{E} | \mu | g} \right|^2}
 = \frac{(q + \epsilon)^2}{1 + \epsilon^2}    \, ,
\end{equation}
where $\epsilon = \frac{E_\text{kin} + E_\text{fin} - E_r}{\pi |V_{Er}|^2}$
(ignoring $F(\underline{E})$, as we do throughout this study)
denotes a reduced energy variable \cite{Fano61}.

The maxima and minima of the purely electronic Fano lineshape are given by
\begin{align}
    \label{eq:Fano_max_el}
    E_{\text{kin,max}} &= E_r + E_\lambda - E_\text{fin}
                          - E_\mu + \frac{\pi |V_{Er}|^2}  q \\
    E_{\text{kin,min}} &= E_r + E_\lambda - E_\text{fin}
                          - E_\mu - q \pi |V_{Er}|^2 \, .
    \label{eq:Fano_min_el}
\end{align}

During the
build-up of the Fano profile,
oscillations in both energy and
time are observed \cite{Fasshauer20_1} originating from interference
effects of the different pathways.

Moreover, the total lifetime of the resonance state can not be extracted
from experimental data by a simple exponential fit, which assumes a
sudden population of the resonance state.
Instead, one needs to take the
population of the resonance state over the entire pulse duration into account.

\subsection{Case 1: 1 vibrational bound state in the electronic resonance state and 1
                    vibrational bound state in the electronic final state}
\label{sec:1l1m}

%\begin{figure}[h]
%\centering
%\includegraphics[width=\columnwidth]{pics/potentials_1l1m.pdf}
%\caption{1l1m}
%\label{fig:pot_1l1m}
%\end{figure}

This case, for which we used the parameters \#1 of Table~\ref{tab:params}
for the simulation,
is the most fundamental one.
Since only one nuclear resonance state and one
nuclear final state are considered,
the derived and used expressions are exact within the limits of the Born-Oppenheimer approximation, the independence of the
electronic bound-continuum coupling matrix element $V_{Er}$ of the
nuclear coordinates and the
purely electronic Fano theory \cite{Fasshauer20_1}.

\begin{figure}[h]
\centering
\includegraphics[width=\columnwidth]{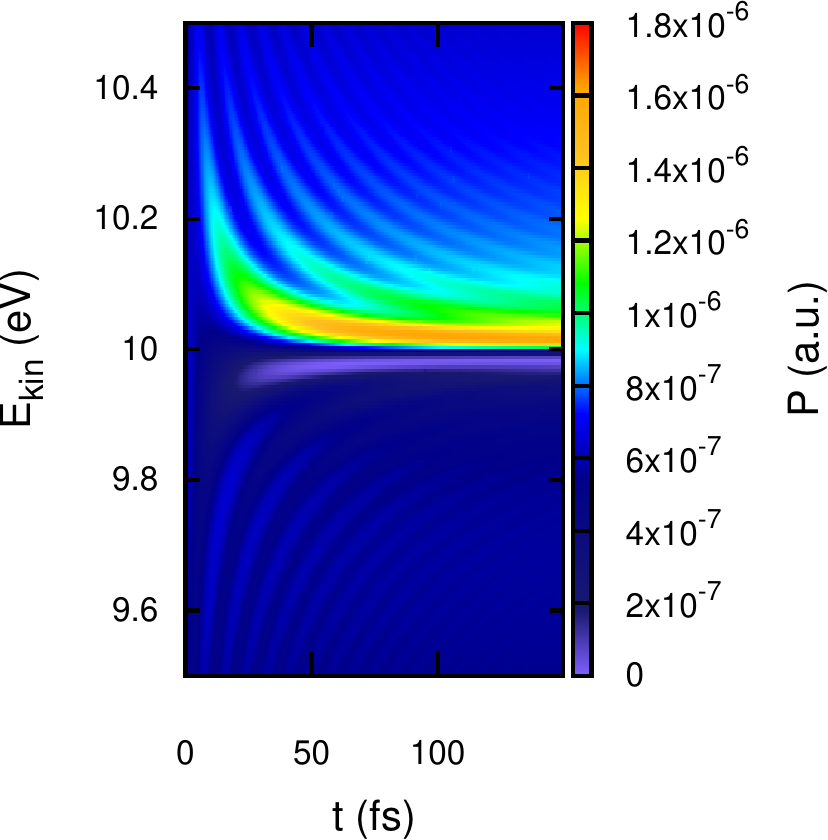}
\caption{(Colour online) Time-resolved kinetic energy spectrum of the RICD electron
for one vibrational state in the electronic resonance state
and one vibrational state in the electronic final state.}
\label{fig:3d_1l1m}
\end{figure}

As to be expected from the purely electronic case, the spectrum
contains a time-resolved build-up of a Fano profile convoluted by the Gaussian caused by the shape of the initiating laser pulse. 

Including the nuclear degrees of freedom changes the details
of the spectrum even in this most simple of cases.
In our description, there are two main reasons for that.

Firstly, the effective decay width $W_\lambda$ of this decay being \qty{21.71}{fs} is
slightly different from the purely electronic one of \qty{20}{fs}.
This deviation from the input lifetime is mainly artificial, though:
According to Eq.~(\ref{eq:Wlambda}),
the decay width is weighted with
$\left|\braket{\chi_{\lambda=0}|\chi_{\mu=0}}\right|^2$, the Franck-Condon
factor between the vibrational resonance state and the vibrational final state.
In this case, its value is \num{0.921}.
But analytical sum rules for Franck-Condon factors %\cite{Nicholls91}
imply that for a given vibrational resonance state $\ket{\chi_\lambda}$,
the sum of its Franck-Condon factors over all vibrational bound final states
plus the integral of its Franck-Condon density over all vibrational continuum final states
must equal unity.
Given that only one bound final state exists in the example system examined in this case,
the above figure of \num{0.921} already represents the bound-states term in the sum rule.
On the other hand, vibrational continuum states are not considered in this study.
This explains the missing \qty{7.9}{\%} of Franck-Condon contributions.
It should be noted, though, that even if vibrational continuum states of the final state
were to be included, such states with an energy above the energy of
the given vibrational resonance state would not contribute to the decay of the latter
because the energy criterion for the RICD process would be violated for these channels.
Thus in each instance, the lifetime of the resonance state will still be increased
with respect to the purely electronic value,
albeit to a much smaller extent than the one yielded by our present simulation.

Secondly, the shape of the signal is altered
compared to the purely electronic one because the relative contributions of the
different pathways are changed by the Franck-Condon overlaps.

%This leads to an effective parameter $q_\text{eff}$, which can be formally defined as
%
%\begin{equation}
% q_\text{eff} = \, \frac{
%    \braket{\chi_{\mu'} | \chi_\lambda} \braket{\chi_\lambda | \chi_0}
% }{
%    \braket{\chi_{\mu'} | \chi_0}
% } \times \frac{
%    \braket{r|\mu|g}
% }{
%    \braket{(\underline{E} - E_{\mu'}) | \mu | g} \pi V^{*}_{(\underline{E} - E_{\mu'}) r}
% } .
%\label{eq:eff_Fano_param}
%\end{equation}
%%\begin{align}
%%\begin{split}
%% &q_{\text{eff}, \lambda} = \, \frac{\braket{\chi_\lambda | \chi_0} \braket{r|\mu|g}}{\pi} \\
%%    &\times \left(
%%        \sum_{\mu'} V^{*}_{(\underline{E} - E_{\mu'}) r}
%%        \braket{\chi_\lambda | \chi_{\mu'}}
%%        \braket{\chi_{\mu'} | \chi_0} \braket{(\underline{E} - E_{\mu'}) | \mu | g}
%%    \right)^{-1} .
%%\label{eq:eff_Fano_param}
%%\end{split}
%%\end{align}
%%so that
%%\begin{align}
%%\begin{split}
%%\label{eq:eff_Fano_shape}
%% &\frac{\left| \braket{\Psi_{\underline{E}} | \mu | G}_\lambda \right|^2}
%% {\left| \sum_{\mu'} \frac{V^{*}_{(\underline{E} - E_{\mu'}) r}}{\sqrt{W_\lambda}}
%%        \braket{\chi_\lambda | \chi_{\mu'}}
%%        \braket{\chi_{\mu'} | \chi_0} \braket{(\underline{E} - E_{\mu'}) | \mu | g} \right|^2}  \\
%% &= \frac{(q_{\text{eff}, \lambda} + \epsilon_\lambda)^2}{1 + \epsilon_\lambda^2}    \, ,
%%\end{split}
%%\end{align}
%%where $\braket{\Psi_{\underline{E}} | \mu | G} = \frac{1}{\sqrt{N_\lambda}} \sum_\lambda \braket{\Psi_{\underline{E}} | \mu | G}_\lambda$.

Both effects,
the Franck-Condon weighting leading to 
different relative contributions of the pathways
and to an effective decay width $W_\lambda$ (the latter being underestimated to some extent
by neglecting continuum vibrational states),
affect the lineshape of the kinetic energy distribution of the emitted electron.
The shape still resembles a Fano profile, though, and we will discuss changes
in peak shapes like the positions of minima and maxima
based on an effective $q$ parameter to make it
comparable to literature in the field. It has to be noted, however, that this
effective $q$ parameter is not rigorously defined.

%The corrected energies of the
%minima and maxima are
%\begin{align}
%    \label{eq:Fano_max}
%    E_{\text{kin,max}} &= E_r + E_\lambda - E_\text{fin}
%                          - E_\mu + \frac{\pi W_\lambda}{q_\text{eff}} \\
%    E_{\text{kin,min}} &= E_r + E_\lambda - E_\text{fin}
%                          - E_\mu - q_\text{eff} \pi W_\lambda
%                          \, .
%    \label{eq:Fano_min}
%\end{align}

For a more narrow ground-state potential and shifted potential energy curves of the
resonance and final states,
the different classes of Franck-Condon overlaps
(ground state - resonance state, ground state - final state and resonance state -
final state) will in most cases be affected differently by alterations
of the potentials in both width and shifts along the internuclear axis.
As a result, the direct, resonance and indirect pathways are weighted
differently, which leads to a change in shape and amplitude of the peak.
%This has the same effect as altering the Fano parameter $q$.
We won't discuss shifting of the potential energy surfaces along the internuclear
axis in detail in the following, but we keep in mind that it can
%causes changes in the effective Fano parameters $q_\text{eff}$ for the 
impact the individual peaks. In extreme cases, decay channels
can become insignificant by shifting the potentials with respect to each other.

\subsection{Case 2: 1 vibrational bound state in the resonance state and 2
                      vibrational bound states in the final state}
\label{sec:1l2m}

In this case,
the parameters \#2 of Table~\ref{tab:params} were used
for the simulation.
Since only the final state has several vibrational bound states, the derived and used
expressions are again exact
within the limits of the Born-Oppenheimer approximation, the independence of the
electronic bound-continuum coupling matrix element $V_{Er}$ of the
nuclear coordinates and the
purely electronic Fano theory.

The total decay width is determined as the sum over the partial decay widths as given by Eq.~\eqref{eq:Wlambda}.
In this case, the effective decay width of \qty{21.45}{fs} again deviates
from the purely electronic one of \qty{20}{fs}
mainly due to incompleteness of the basis set of vibrational eigenfunctions.
The sum of the Franck-Condon factors for the vibrational resonance state
over both vibrational bound final states equals \num{0.941}.
The distance to unity is therefore smaller than in the previous case,
which means that by ignoring the vibrational continuum states of the final state,
a lesser error in the effective decay width is introduced in the present case.

%\begin{figure}[h]
%\centering
%\includegraphics[width=\columnwidth]{pics/potentials_1l2m.pdf}
%\caption{1l2m}
%\label{fig:pot_1l2m}
%\end{figure}

\begin{figure}[h]
\centering
\includegraphics[width=\columnwidth]{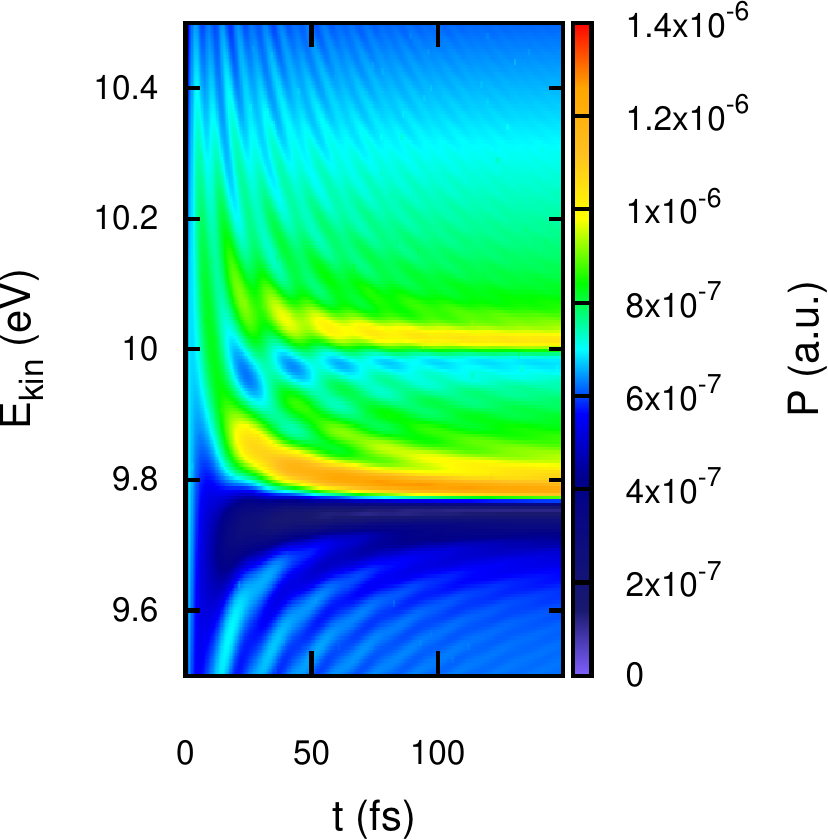}
\caption{(Colour online)
Time-resolved RICD spectrum for one vibrational state in the electronic resonance state and two different vibrational states in the electronic final state. Two peaks are visible, one for each vibrational state in the final state. The peak at \qty{10}{eV} corresponds to the vibrational $0\leftarrow 0$ transition, while the one at \qty{9.8}{eV} corresponds to the $1\leftarrow 0$ transition. Both peaks exhibit the characteristic Fano profile. For a detailed discussion, see the main text.}
\label{fig:3d_1l2m}
\end{figure}

The corresponding time-resolved spectrum
shown in Figure~\ref{fig:3d_1l2m} has two
Fano peaks with maxima at \qty{10.015}{eV} and \qty{9.785}{eV} corresponding
to the transitions $\mu=0 \leftarrow \lambda=0$ and $\mu=1 \leftarrow \lambda=0$,
respectively.

We notice that the intensities of the two peaks differ
and the overall spectrum is more complex than in case 1.

The corresponding Franck-Condon overlaps then determine the boost of the
different decay peaks.
In this particular case, they are
$\braket{\chi_{\mu=0}|\chi_0} = 0.38$ and
$\braket{\chi_{\mu=1}|\chi_0} = 0.82$.
Consequently, the direct contribution for the final state with $\ket{\chi_{\mu=1}}$ is higher than that for the
vibrational ground state.

The direct parts of the transition amplitudes also affect both the relative
peak intensity and the peak
shape of the corresponding decay peaks by changing the relative contributions
of the direct, the resonance and the indirect terms. The other contributions
are affected by the Frank-Condon overlaps involving the resonance state. In this
example, they are
$\braket{\chi_\lambda|\chi_0} = 0.87$,
$\braket{\chi_{\mu=0}|\chi_\lambda} = 0.55$ and
$\braket{\chi_{\mu=1}|\chi_\lambda} = 0.80$.
Since $\braket{\chi_{\mu=1}|\chi_\lambda}$ is larger than
$\braket{\chi_{\mu=0}|\chi_\lambda}$, the transition from the resonance state
to the vibrational excited final state is favoured.

In order to determine the relative contributions of the different pathways to
the respective signal, we need to compare $\braket{\chi_{\mu}|\chi_0}$ to
the product $\braket{\chi_{\mu}|\chi_\lambda} \braket{\chi_\lambda|\chi_0}$
according to Eq.~(\ref{eq:ampl_working_eq}). These products are 0.48 ($\mu=0$) and 
0.69 ($\mu = 1$). Hence, for the vibrational ground state of the electronic final state with $\mu = 0$,
the relative contribution of the resonance
pathway is increased, while it is decreased for the vibrational excited state
with $\mu = 1$. Qualitatively, this leads to the same effect as increasing an
effective Fano parameter for the
vibrational ground state of the final state and decreasing such an
effective Fano parameter for the
vibrational excited state.
%Since the extrema of a Fano shaped peak depend on the Fano parameter as
%shown in Eqs.~(\ref{eq:Fano_max}, \ref{eq:Fano_min}),
%\begin{align}
%    E_{\text{kin,max}} &= E_r + E_\lambda - E_\text{fin}
%                          - E_\mu + \frac{\pi W_\lambda}q \\
%    E_{\text{kin,min}} &= E_r + E_\lambda - E_\text{fin}
%                          - E_\mu - q \pi W_\lambda
%\end{align}
The minima and maxima of the peaks are shifted in energy with
respect to a purely electronic $q$ parameter.
The maximum and minimum of the peak corresponding to
the decay into the vibrational ground state of the electronic
final state are shifted towards lower kinetic energies with respect to the energies expected for the
purely electronic $q$ parameter,
while they are shifted towards higher energies for the peak
corresponding to the decay to the vibrational excited state of
the electronic final state.

%\begin{figure}[h]
%\centering
%\includegraphics[width=\columnwidth]{pics/time_1l2m_res.pdf}
%\caption{1l2m}
%\label{fig:time_1l2m}
%\end{figure}

During build-up of the signals, the interference-induced oscillations in
time and energy of the two peaks overlap to yield a complex pattern.
Despite the resulting additional oscillation pattern in time, this is not due to an
additional interference effect, because the different final-state
contributions are, according to Eq.~(\ref{eq:td_probability}), simply added up
to yield the total spectrum.
%(That is in contrast to Case~4
%where two competing resonance pathways into the same final state
%do exist and interfere.)

A Fourier analysis of the oscillations in time for an exemplary kinetic energy
of \qty{9.6}{eV} shows two peaks with periods of
\qty{25}{fs} and \qty{10}{fs}. They correspond to the interference
between the resonance and the two direct pathways
as already expected from the study of the purely electronic case
\cite{Fasshauer20_1}.

\subsection{Case 3: 1 vibrational bound state in the resonance state and several
                      vibrational bound states in the final state}
\label{sec:1lmanym}

\begin{figure}[h]
\centering
\includegraphics[width=\columnwidth]{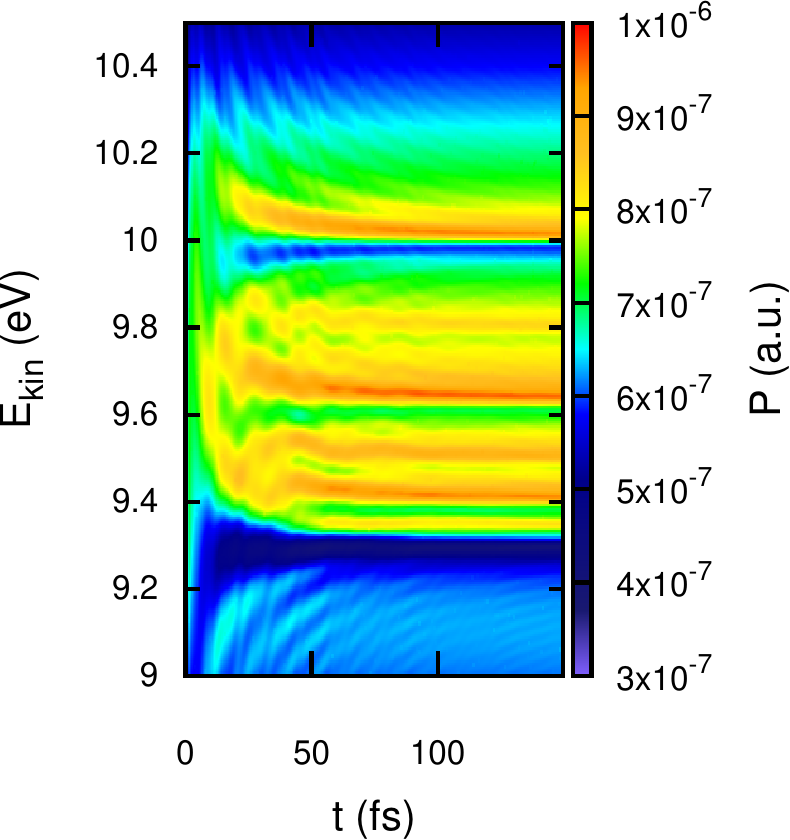}
\caption{(Colour online) Time-resolved RICD spectrum for one vibrational state in
         the electronic resonance state and seven vibrational
         states in the electronic final state.}
\label{fig:3d_1lmanym}
\end{figure}

In this case, 
the parameters \#3 of Table~\ref{tab:params} were used
for the simulation of the time-resolved RICD spectrum shown
in Figure~\ref{fig:3d_1lmanym}.
The final state has seven vibrational bound states with energies between
\qty{0.1158}{eV} and \qty{0.8090}{eV} with respect to the
electronic minimum energy.
As for case 2, only the final state has several vibrational bound states,
and therefore, the exactness is only limited by the same
assumptions as in cases 1 and 2.

Analogous to case 2, the overall spectrum is
broadened towards lower kinetic energies
by the possibility of ending in multiple vibrational final states.
Moreover, the individual peak intensities and shapes are affected by the
different Franck-Condon overlaps in the same way as in case 2.
This is also reflected in the effective decay width of this case following
Eq.~\eqref{eq:Wlambda}, which corresponds to \qty{20.28}{fs}, although
the deviation with respect to the purely electronic case is, again, mainly
due to the neglected contributions of vibrational continuum states.
Consequently, the manifold of signals yields a spectrum in which the oscillations in
energy of the individual peaks become smeared out throughout the spectrum.

\begin{figure}[h]
\centering
 \includegraphics[width=\columnwidth]{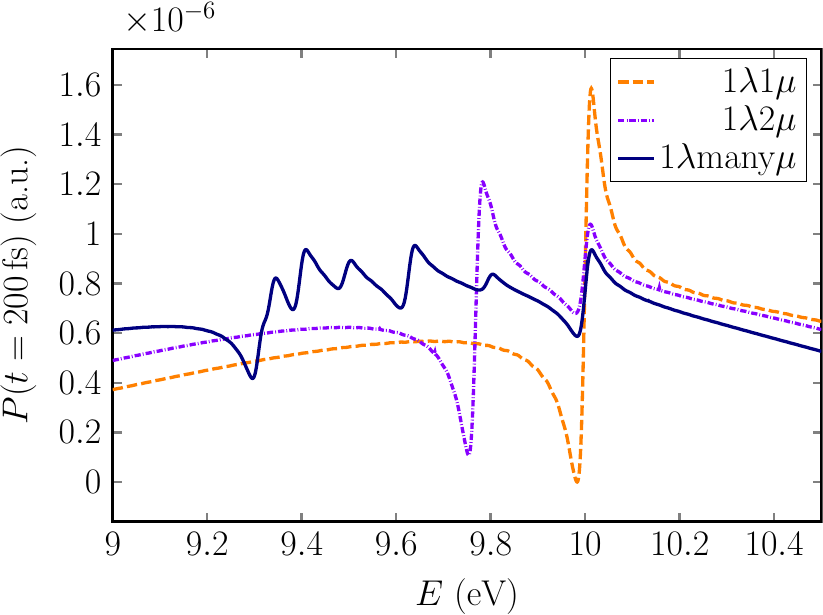}
\caption{(Colour online) RICD spectra at time $t=\qty{200}{fs}$ 
    for cases 1 (dashed curve in orange), 2 (dash-dotted curve in violet) 
    and 3 (solid curve in dark blue).}
\label{fig:1lxmE}
\end{figure}

Despite of having seven unique vibrational states in the electronic final
state, only six clearly separated positive signals are visible. The
two highest vibrational excited states are too close to be resolved and
are instead combined into one peak as shown 
in the orange curve of Figure~\ref{fig:1lxmE}
at around \qty{9.3}{eV}.

\subsection{Case 4: 2 vibrational bound states in the resonance state and 1
                      vibrational bound state in the final state}
\label{sec:2l1m}

This case provides the opportunity to discuss several
features, amongst them the coherence between the two
vibrational states in the resonance state. In order to give
a thorough analysis, we have therefore chosen to display
different sets of parameters.

\subsubsection{Large energy splitting between the vibrational
               resonance states, $q=1$}
\label{sec:2l1m_q1}
In this case,
the parameters \#4 of Table~\ref{tab:params} were used
for the simulation.
Here, the potential energy curve of the resonance state supports two vibrational bound
states, which are split by \qty{0.241}{eV}. The effective lifetimes of these
two states are \qty{111.93}{fs} and \qty{37.67}{fs} for the vibrational ground
and excited state, respectively. These are significantly larger than the input
values for the purely electronic lifetime of \qty{20}{fs}
each.
This may indicate a serious incompleteness of the vibrational
eigenfunctions basis set, especially in the case of the vibrational ground
state of the resonance state. However as of yet, we did not quantify to
which degree this prolongation of lifetimes is actually physical due to
decay channels through high-lying states being energetically inaccessible.

%\begin{figure}[h]
%\centering
%\includegraphics[width=\columnwidth]{pics/potentials_2l1m.pdf}
%\caption{2l1m}
%\label{fig:pot_2l1m}
%\end{figure}

\begin{figure}[h]
\centering
\includegraphics[width=\columnwidth]{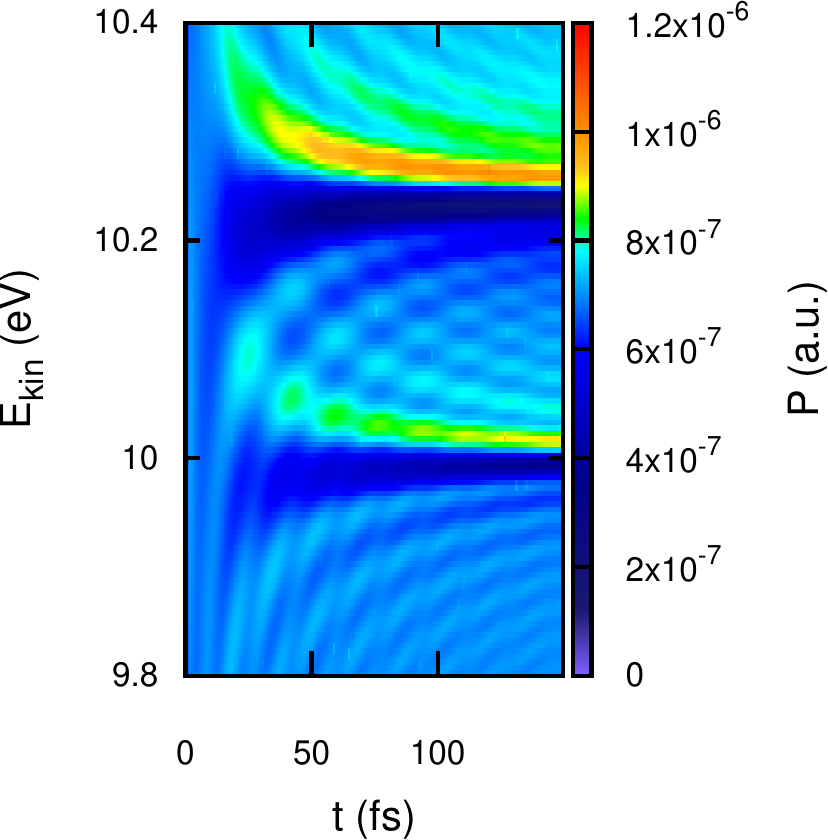}
\caption{(Colour online) Time-resolved RICD spectrum for two vibrational states in the electronic resonance state and a single vibrational state in the electronic final state. The peak at \qty{10}{eV} corresponds to the vibrational $0\leftarrow 0$ transition, while the one at \qty{10.2}{eV} corresponds to the $0\leftarrow 1$ transition. Both peaks again exhibit the characteristic Fano profile, plus an additional interference pattern created by the additional pathway via the vibrational excited state in the resonance state. For a detailed discussion, see the main text.}
%\caption{Time-resolved kinetic energy spectrum of the RICD electron for two vibrational bound states in the electronic resonance state and a single vibrational state in the electronic final state. Two peaks are visible, one for each vibrational state in the resonance state. The peak at \qty{10}{eV} corresponds to the vibrational $0\leftarrow 0$ transition, while the one at \qty{10.2}{eV} corresponds to the $0\leftarrow 1$ transition. Both peaks exhibit the characteristic Fano profile plus an additional interference pattern created by the additional pathway via the vibrational excited state in the resonance state. For a detailed discussion, see the main text.}
\label{fig:3d_2l1m}
\end{figure}

The simulation with these parameters yields the spectrum shown in Figure
\ref{fig:3d_2l1m}. It is characterized by two peaks, corresponding to the
vibrational states of the electronic resonance state.
Contrary to cases 2 and 3,
the signal corresponding to the vibrational excited state of the resonance state
has a higher kinetic energy than the one corresponding to its vibrational
ground state.
Furthermore, the spectrum is not broadened by the possibility of exciting into
several vibrational final states in the indirect pathway and, hence, the peak of the vibrational excited
state of the electronic resonance state is slightly
disfavoured by the energy profile of the laser pulse.

Again, the Franck-Condon parameters influence the peak intensity and shape,
and since
$\braket{\chi_{\mu}|\chi_0} = 0.91$ is significantly larger than both the
products of 
$\braket{\chi_{\lambda=0}|\chi_0} = 0.38$
with $\braket{\chi_{\mu}|\chi_{\lambda=0}} = 0.42$
and 
$\braket{\chi_{\lambda=1}|\chi_0} = 0.63$ with
$\braket{\chi_{\mu}|\chi_{\lambda=1}} = 0.73$, the effective values of the
Fano parameter are decreased. This favours the direct pathway relative to the pathway via the resonance state.
Amongst the resonance pathways, the signal corresponding to the route via the vibrational
excited state of the resonance state
is significantly favoured over the one via its vibrational ground state.

The overall spectrum has a complex pattern created by the
interference of both the direct and resonance pathways.
Moreover, in this case, it should also be possible to
observe the interference between the pathways via the two
different vibrational resonance states through
an oscillation in time as per
\begin{equation}
    \label{eq:res_res_int}
    P \propto \cos\left[ (E_{\lambda=1} - E_{\lambda=0}) \, t \right]   \; .
\end{equation}

\begin{figure}[h]
\centering
\includegraphics[width=\columnwidth]{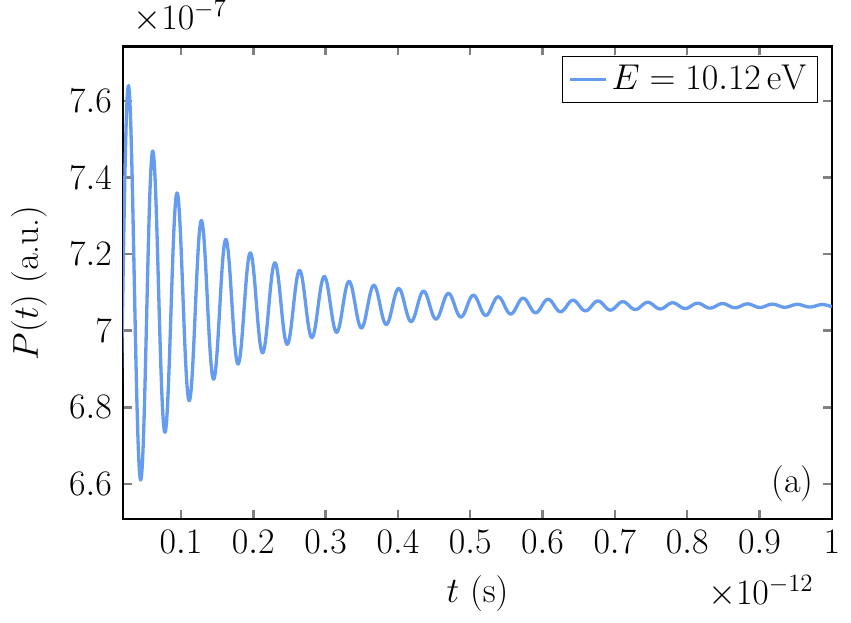}
\includegraphics[width=\columnwidth]{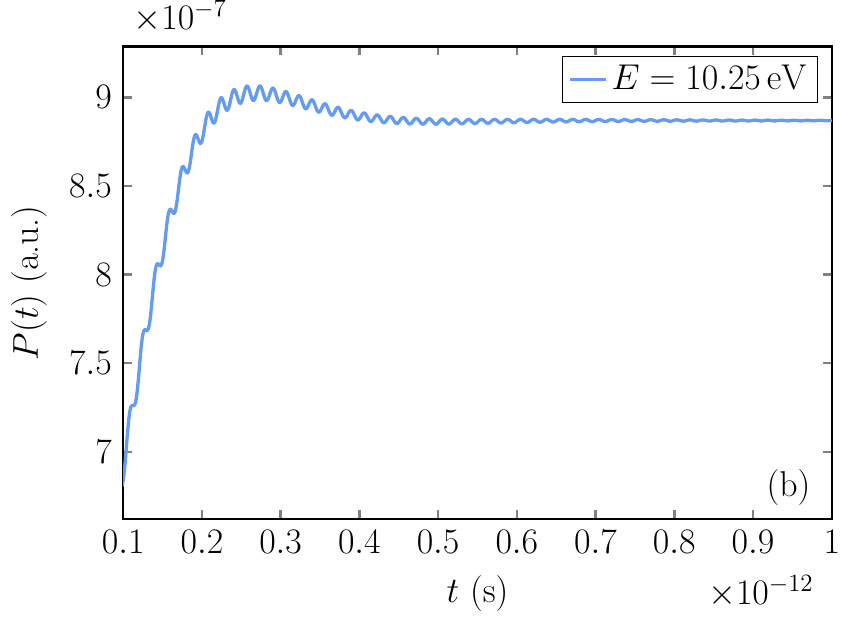}
\caption{Profiles in time for two vibrational states within the resonance state
    and one vibrational state within the final state (Figure~\ref{fig:3d_2l1m}).
    They have been evaluated at kinetic energies of \textbf{a)} \qty{10.12}{eV} and \textbf{b)} \qty{10.25}{eV}.
    A few data points which were affected by numerical noise have been discarded.}
\label{fig:osc_2l1m}
\end{figure}

Figure~\ref{fig:osc_2l1m} allows for a more detailed analysis of the interference pattern. It
shows the oscillations in time at energies \textbf{a)} \qty{10.12}{eV}, which
is chosen to include the minima and maxima of the observed
interference pattern of the spectrum, and
\textbf{b)} \qty{10.25}{eV}, which is close to the maximum of the peak
corresponding to the pathway via the
vibrational excited state of the electronic resonance state.

From earlier studies of time-resolved Fano profiles \cite{Wickenhauser05,Kaldun16,Gruson16,Fasshauer20_1},
we know that for one resonance state, the
peak first builds up and then decreases exponentially according with the electronic
decay. In Ref.~\cite{Fasshauer20_1}, we showed that the interference between
the direct pathway and the resonance pathway additionally leads to an oscillation
in time, whose frequency is determined by the energy difference between the
investigated energy of the cut through the spectrum and the theoretical
resonance position.

For the case of two resonance states, irrespective of whether it is an electronic or
a vibrational state, the interference between the contributions of the
two resonance pathways to the wavefunction
leads to another
oscillation, whose period is determined by the energy difference between the
two resonance states.
Eq.~(\ref{eq:res_res_int}) is an instance for such an oscillation.
For the energy difference of this example, one would
expect one cycle to last for \qty{17}{fs}.

In Figure~\ref{fig:osc_2l1m} \textbf{a)}, the cut between the peaks
shows a clear oscillation in time
with decreasing intensity. Its full period is \qty{33.6}{fs} and hence
twice the period one would expect for the resonance-resonance interference pattern.
In Figure~\ref{fig:osc_2l1m} \textbf{b)} with the cut close to the resonance,
build-up occurs according to earlier studies. Overlaying this feature,
an oscillation
is visible which after ca. \qty{200}{fs} has a stable period of
\qty{16.3}{fs}, close to the above theoretical value of \qty{17}{fs}.

In neither case, the oscillation is caused by the resonance-resonance interference
but rather by the interference of the direct with the respective resonance pathways.
For the case of Figure~\ref{fig:osc_2l1m} \textbf{a)},
considering the kinetic energy of \qty{10.12}{eV}, the  
energy difference to the kinetic energy corresponding to the decay
from the vibrational ground state of the resonance state is given by
\qty{0.12}{eV} corresponding to an
oscillation period of \qty{34.5}{fs}. This is in good agreement with
the observed \qty{33.6}{fs}.
The relationship with the
period of the resonance-resonance interference is only coincidental.

For the case of Figure~\ref{fig:osc_2l1m} \textbf{b)} with a kinetic energy of \qty{10.25}{eV}, the
observed period
almost equals the expected resonance-resonance interference period. However, at this kinetic energy, the period
of the oscillation caused by the interference between the direct and the resonance
parts of the neighbouring peak are the same as the period of the resonance-resonance
interference. It is therefore not possible to distinguish between the two close to a
resonance energy.

\subsubsection{Large energy splitting between the vibrational
               resonance states, $q=10$}
\label{sec:2l1m_q10}

In section \ref{sec:2l1m_q1}, the contributions of the resonance pathways
were comparably small.
Whether the resonance-resonance interference is strong enough to have an
impact on the overall signal depends on both the intrinsic property of the
$q$ parameter --- the larger $q$, the larger the resonance contribution ---
and the energetic width of the resonance part, which is inversely
proportional to the temporal width and therefore also the number of cycles
of the exciting pulse. It can hence
be manipulated: the shorter the pulse, the larger the resonance signal
at kinetic energies different from the peak maximum. This can directly
be seen by inspection of the second part of Eq.~(\ref{eq:res_ampl}).
In order to highlight the resonance-resonance interference and its
properties, we increase the purely electronic Fano parameter $q$ from 1
to 10.

\begin{figure}[h]
\centering
\includegraphics[width=\columnwidth]{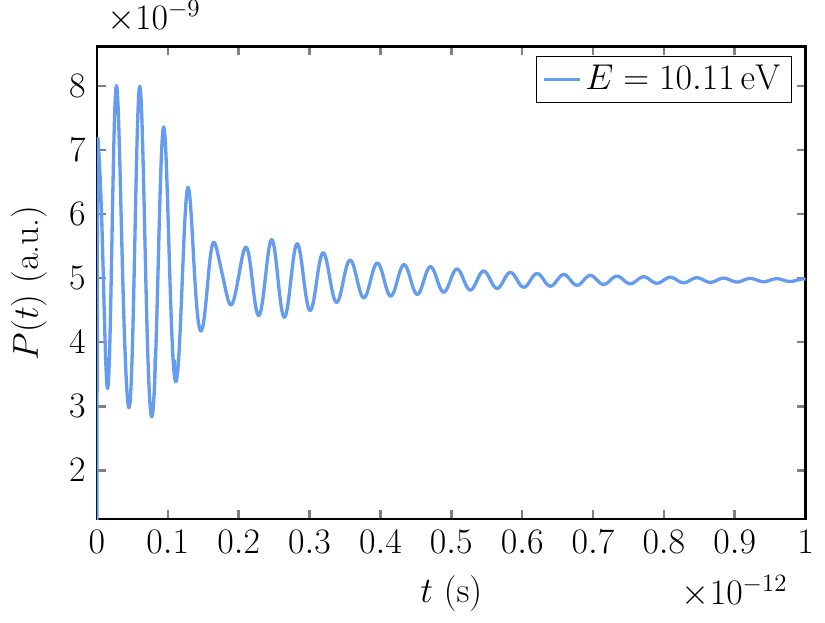}
\caption{Profile in time for two vibrational states within the resonance state 
    and one vibrational state within the final state (similar to Figure~\ref{fig:3d_2l1m})
    for $q=10$, evaluated at a kinetic energy of \qty{10.11}{eV}.
    A few data points which were affected by numerical noise have been discarded.}
\label{fig:osc_2l1m_q10}
\end{figure}

The resulting time-resolved spectrum is very similar to that for
$q=1$ in Fig.~\ref{fig:3d_2l1m}. The Fourier analysis of the oscillation
pattern evaluated at a kinetic energy of \qty{10.11}{eV} and shown in
Fig.~\ref{fig:osc_2l1m_q10} is, however, significantly
different. It shows three different oscillations with periods \qty{36.9}{fs},
\qty{31.9}{fs} and \qty{17.7}{fs}. The first two periods correspond to the two
interferences of the two different vibrational resonance pathways with the
direct pathway. The one with a period of \qty{17.7}{fs} however corresponds
to the resonance-resonance interference, which from the energy
difference is expected to have a period of \qty{17.2}{fs}.

\subsubsection{Small energy splitting between the vibrational
               resonance states, $q=10$}

\begin{figure}[h]
\centering
\includegraphics[width=\columnwidth]{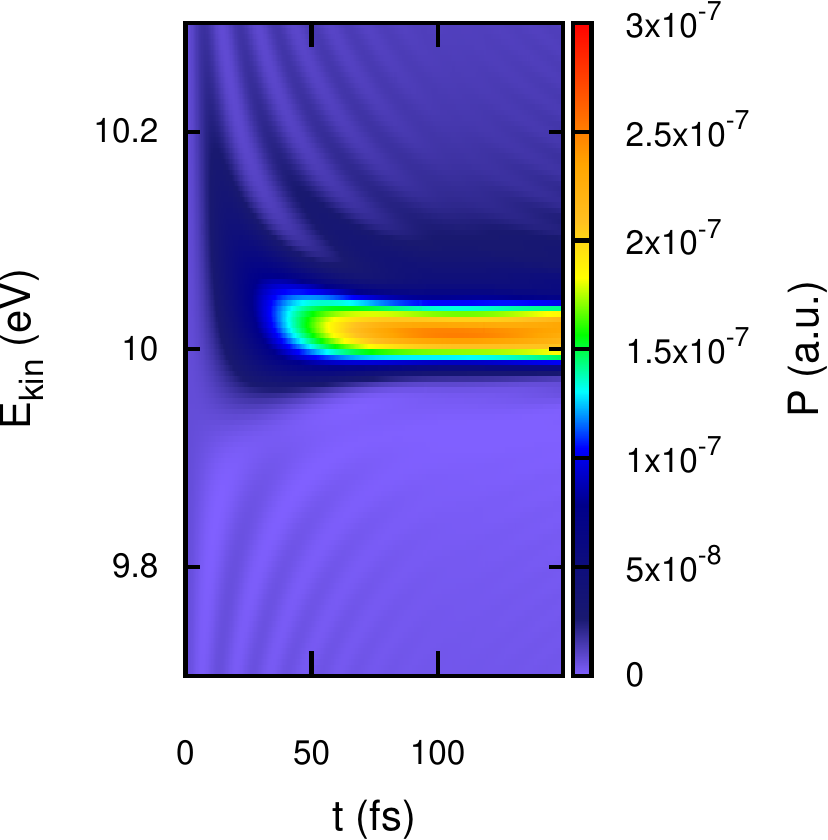}
\caption{(Colour online) Time-resolved RICD spectrum for two energetically close vibrational states in the electronic resonance state and a single vibrational state in the electronic final state for $q=10$.}
\label{fig:full_2l1m_smalldE}
\end{figure}

Until now, we have seen scenarios in which the vibrational resonance
states were well separated in energy. However, this is not necessarily the case,
and even most likely not.
Using the parameters \#5 given in Table~\ref{tab:params} yields
an energy difference for the vibrational resonance states of \qty{0.0191}{eV}.
As in section \ref{sec:2l1m_q10} we have chosen a larger Fano parameter
of $q=10$ in order to both narrow the individual peaks and
increase the contribution of the resonance pathways.
Nevertheless, the corresponding peaks in the spectrum cannot be resolved, as shown in Figure
\ref{fig:full_2l1m_smalldE}.

However, if the
contributions of the resonance parts of the wavefunction are substantial,
the resonance-resonance interference
can in principle be observed at energies that are different from the
kinetic energy corresponding to the resonance
and hence be distinguished from direct-resonance interferences.
Furthermore, from the oscillation
period of the signal, the energy difference between the two resonance
states can be determined.

\begin{figure}[h]
\centering
\includegraphics[width=\columnwidth]{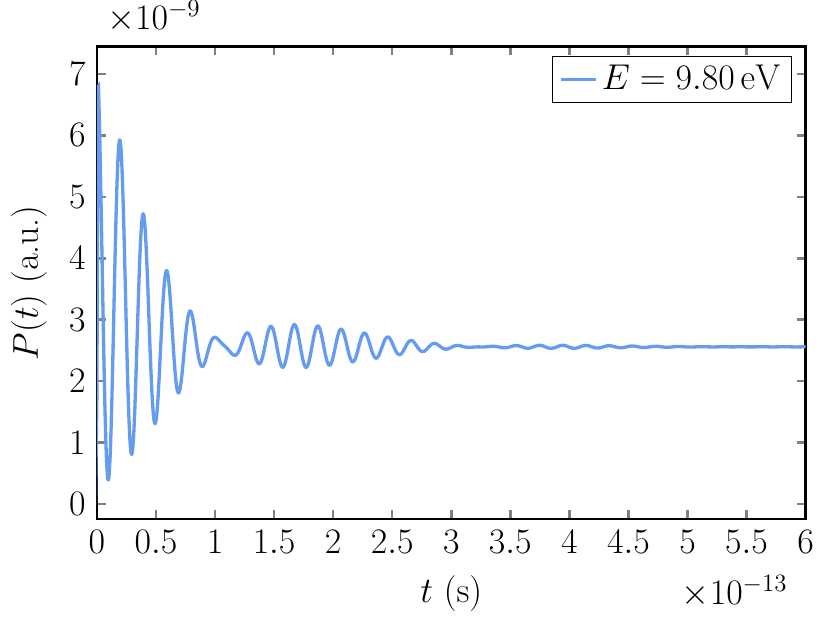}
\caption{Profile in time for two energetically close vibrational states within the resonance state 
    and one vibrational state within the final state for $q=10$ (Figure~\ref{fig:full_2l1m_smalldE}), 
    evaluated at a kinetic energy of \qty{9.80}{eV}.
    A few data points which were affected by numerical noise have been discarded.}
\label{fig:osc_2l1m_smalldE}
\end{figure}

In Figure~\ref{fig:osc_2l1m_smalldE}, we give an example for such an
oscillatory variation in signal intensity for the parameter set \#5
at a kinetic energy of \qty{9.8}{eV}. It shows two oscillations.
On the one hand, the oscillation with a period of \qty{21}{fs} is caused
by the interference of the direct
term with either of the resonance ones. Since they are energetically so similar,
the difference in the oscillation periods is not visible.
It is only in the Fourier-transformed signal that a double peak can be seen,
corresponding to oscillation periods of \qty{18.7}{fs} and \qty{20.9}{fs}, respectively.
On the other hand, the oscillation with the longer period of ca. \qty{220}{fs}
is caused by the resonance-resonance interference.

We note that it may be experimentally difficult to resolve these variations
in signal intensities in low intensity areas.

\subsection{Case 5: 2 vibrational bound states in the resonance state and 2
                      vibrational bound states in the final state}
\label{sec:2l2m}

\subsubsection{Large energy splitting between the vibrational
               resonance states}
\label{sec:2l2m_largeres}

In this case,
the parameters \#6 of Table~\ref{tab:params} were used
for the simulation.
Both the resonance and the final state
support two vibrational bound states. This is the smallest possible case in which
all hitherto discussed properties are combined.
The energy differences of the vibrational states are \qty{0.2434}{eV} for the
resonance state and \qty{0.0493}{eV} for the final state.

%\begin{figure}[h]
%\centering
%\includegraphics[width=\columnwidth]{pics/potentials_2l2m.pdf}
%% currently reverse
%\caption{2l2m}
%\label{fig:pot_2l2m}
%\end{figure}

\begin{figure}[h]
\centering
\includegraphics[width=\columnwidth]{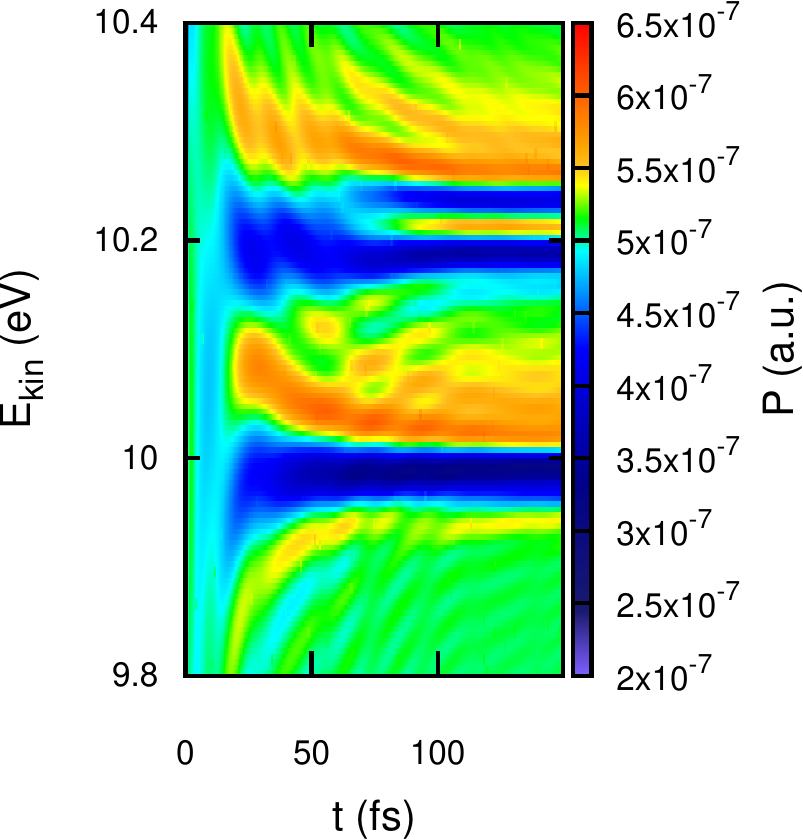}
\caption{(Colour online) Time-resolved RICD spectrum for two vibrational states
         in both the electronic resonance and final state with
         a large energy difference between the vibrational
         resonance states and a small energy difference between the vibrational final states.}
\label{fig:3d_2l2m}
\end{figure}

These parameters lead to the spectrum shown in Figure~\ref{fig:3d_2l2m}. It
is characterized by two large peaks, which are split into the two peaks for the
different vibrational final states.

Since the energy difference between the two vibrational final states is very small,
the width of the overall peak is hardly affected by their presence.

The Franck-Condon overlaps for the transition between the ground and the final
state relevant for the direct pathway are
$\braket{\chi_{\mu=0}|\chi_0} = 0.60$ and
$\braket{\chi_{\mu=1}|\chi_0} = 0.49$, which indicates that the direct
pathways contribute with similar weights. The Franck-Condon overlaps relevant
for the decay from the resonance states are more interesting.
They are
$\braket{\chi_{\lambda=0}|\chi_0} = 0.40$,
$\braket{\chi_{\lambda=1}|\chi_0} = 0.33$,
$\braket{\chi_{\mu=0}|\chi_{\lambda=0}} = 0.87$,
$\braket{\chi_{\mu=1}|\chi_{\lambda=0}} = -0.27$,
$\braket{\chi_{\mu=0}|\chi_{\lambda=1}} = 0.37$,
and
$\braket{\chi_{\mu=1}|\chi_{\lambda=1}} = 0.57$.
Hence,
%the effective Fano parameters $q_\text{eff}$
%according to Eq.~(\ref{eq:eff_Fano_param})
%are all decreased and 
the peak intensities and shapes are affected,
in accordance with decreased effective Fano $q$ parameters.
Moreover, $\braket{\chi_{\mu=1}|\chi_{\lambda=0}}$, the Franck-Condon overlap between
the vibrational ground state of the resonance state and
the vibrational excited state of the final state,
is negative and so is the corresponding effective
Fano parameter. % $q_\text{eff}$.
Under sign change of the effective $q$ parameter, the peak shape becomes its own mirror
image, which therefore explains the unexpected behaviour for the peak at
\qty{9.95}{eV}.
In this particular example, its minimum coincides with the minimum of the
neighbouring peak and the typical Fano lineshape becomes more difficult to identify.

At a kinetic energy of \qty{10.02}{eV}, an interference pattern is present
that includes an oscillation with an apparent period of ca. \qty{21}{fs}.
Fourier analysis reveals this to be in fact two oscillations
with periods of \qty{19}{fs} and \qty{24}{fs}, respectively,
which are too close for the two oscillations  to be distinguishable in the time-resolved signal.
The energy difference between the vibrational resonance states relates to an oscillation with
a \qty{17}{fs} period, which fits nicely. However, the signals corresponding to
direct-resonance interference that involves
the vibrational excited state of the resonance state indicate oscillations with
periods of \qty{19}{fs} and \qty{24}{fs} for $\mu=0,1$, respectively.
An unambiguous attribution of the observed oscillations
to the underlying pathways that give rise to them is therefore impeded:
The interference pattern is obviously more complicated than
in case 4 due to the
manifold of different contributions for any chosen kinetic energy.

\subsubsection{Small energy splitting between the vibrational
               resonance states}
\label{sec:2l2m_smallres}

\begin{figure}[h]
\centering
\includegraphics[width=\columnwidth]{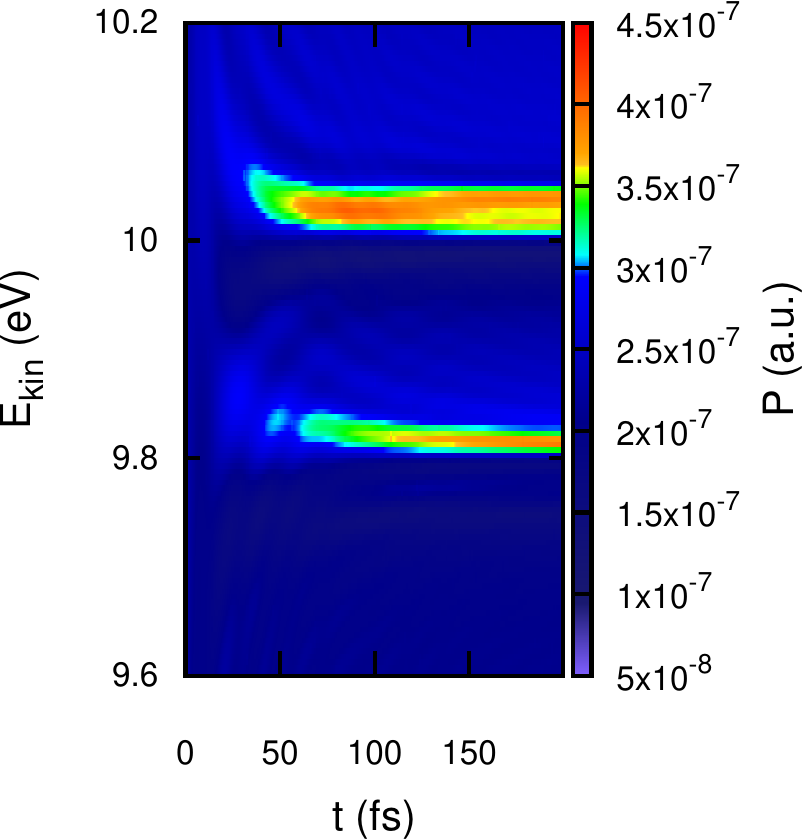}
\caption{(Colour online) Time-resolved RICD spectrum for two vibrational states
         in both the electronic resonance and final state with
         a small energy difference between the vibrational
         resonance states and a large energy difference between the vibrational final states.}
\label{fig:3d_2l2m_reverse}
\end{figure}

When the parameters of the resonance- and the final-state potentials are interchanged
(parameters \#7 of Table~\ref{tab:params}),
the energy difference between the two resonance states becomes very small and the
two larger signals are clearly separated from another as illustrated
in Figure~\ref{fig:3d_2l2m_reverse}. This may allow for an
easier interpretation of the interference spectrum. 

Compared to the spectrum in Figure~\ref{fig:3d_2l2m}, the different direct pathways
boost only those parts of the spectrum where the peaks are.
Due to the Franck-Condon overlaps of this example, one part of the
double peak around \qty{9.8}{eV}
is hardly visible, and we will therefore focus on the other double peak.
As in the first example of this case, the negative sign of
$\braket{\chi_{\mu=0}|\chi_{\lambda=1}}$, the Franck-Condon overlap between
the vibrational excited state of the resonance state and
the vibrational ground state of the final state,
leads to an unusual line shape for the peak around \qty{10.04}{eV}.
However, there is also a clear contrast to the previous example, in that
the small energy difference between the two vibrational states
of the resonance state and their
approximately even contributions to the spectrum in the double
peak around \qty{10}{eV}
should, according to Eqs.~(\ref{eq:ampl_working_eq}) and (\ref{eq:res_res_int}),
lead to a clear oscillation pattern in time
caused by the resonance-resonance interference with
comparatively long periods.

\begin{figure}[h]
\centering
\includegraphics[width=\columnwidth]{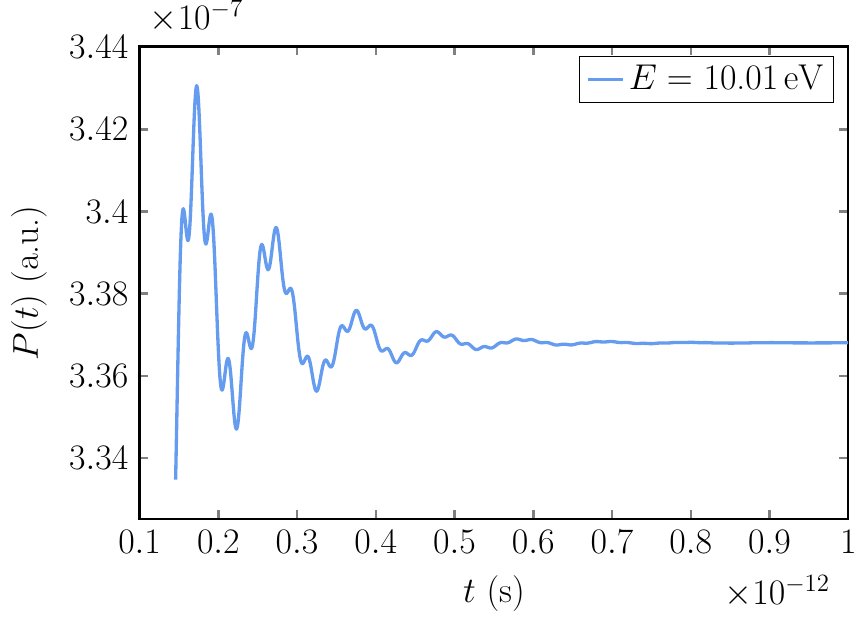}
\caption{Profile in time
    for two energetically close vibrational states within the resonance state
    and two energetically more distant vibrational states in the final state
    (Figure~\ref{fig:3d_2l2m_reverse}), evaluated at a kinetic energy of \qty{10.01}{eV}.
    A few data points which were affected by numerical noise have been discarded.}
% \caption{Profile in time
% for two vibrational states within both the electronic resonance state and final state
% with a small energy difference between the vibrational resonance states
% and a large energy difference between the vibrational final states
% (Figure~\ref{fig:3d_2l2m_reverse}), evaluated at a kinetic energy of \qty{10.01}{eV}.}
\label{fig:time_2l2m_reverse}
\end{figure}

Due to
this long oscillation period,
the resonance-resonance
interference should already close to the peak maxima be distinguishable
from the interference with the direct pathway. This allows an evaluation with
more significant peak intensity.
We therefore focus on the time-resolved pattern for a kinetic energy of
\qty{10.01}{eV} close to the
$0\leftarrow 0$ peak, which is shown in Figure~\ref{fig:time_2l2m_reverse} after
the first \qty{145}{fs}.
Upon first inspection, the signal can be described by two overlapping oscillations
with very different frequencies, whose amplitude is decreased exponentially.
A Fourier analysis of the signal shows that
the dominant and the perturbing oscillation have periods of
\qty{100}{fs} and only \qty{20}{fs}, respectively.
% We therefore fitted the data to the following function:

% \begin{equation}
%  f_t(t) = a \cos\Bigl[ \frac tb +c \Bigr] exp\Bigl[ -\frac{t-d}{2e} \Bigr]
%           + f \exp \Bigl[ -(t-g) / 2h \Bigr] + i
% \end{equation}

% and show it as the dotted line in Figure~\ref{fig:time_2l2m_reverse}.
% The dominant oscillation has a period of ca. \qty{100}{fs}, which is increasing
% over time, while the perturbing oscillation is highly irregular with periods
% fluctuating between \qty{15}{fs} and \qty{26}{fs}.

According to the absolute square of Eq.~(\ref{eq:ampl_working_eq}), five different kinds
of oscillations could in principle be observable involving all possible combinations
of vibrational states in the electronic resonance $\lambda=0,1$ and final state
$\mu=0,1$ as well as
the resonance-resonance contribution from the two different vibrational states
$\lambda=0,1$.
%\begin{align}
% P(t) &\propto
% \cos \Bigl[ t (E_{\lambda=0} - E_{\lambda=1}) \Bigr]
% \exp \Bigl[ -\pi (W_0 + W_1) t \Bigr]   \\
% P(t) &\propto
% \cos\Bigl[ t (E_\text{kin} + E_\text{fin} + E_\mu - E_r - E_\lambda) \Bigr]
% \exp\Bigl[ -\pi W_\lambda t \Bigr],
%\end{align}
%
%where $\lambda=0,1$ and $\mu=0,1$.

The corresponding theoretical periods are given in
Table~\ref{tab:periods}.

\begin{table}[h]
 \caption{Oscillation periods for the different interference combinations.}
 \label{tab:periods}
 \begin{tabular}{p{0.2\columnwidth}p{0.5\columnwidth}r}
  \toprule
  \# & characteristics                   & $T \, [\mathrm{fs}]$ \\
  \midrule
   1 & $\mu=0 , \lambda=0$     & 414 \\
   2 & $\mu=1 , \lambda=0$     &  16 \\
   3 & $\mu=0 , \lambda=1$     & 105 \\
   4 & $\mu=1 , \lambda=1$     &  20 \\
   \midrule
   5 & $\lambda=0, \lambda=1$ &  84 \\
  \bottomrule
 \end{tabular}
\end{table}

The observed low-frequency oscillation is most likely inflicted by the
interference between contributions from the
vibrational excited resonance state and the vibrational ground state
(parameters \#3 of Table~\ref{tab:periods}), which also has a non-negligible
contribution at \qty{10.01}{eV}. The observed high-frequency oscillation can
predominantly be attributed to the interference between the
contributions from the two vibrational excited states
(\#4 of Table~\ref{tab:periods}), because
the contributions from the pathway involving the vibrational ground state of
the resonance state are disfavoured by the Franck-Condon overlaps.
The oscillation caused by the interference of the two resonance states decaying to
the vibrational ground state is not visible, because the other contributions
cover them.

%%%%%%%%%%%%%%%%%%%%%%%%%%%%%%%%%%%%%%%%%%%%%%%%%%%%%%%%%%%%%%%%%%%%%%%%%%%%%%%%%
%\input{subsecs/approximations}

\section{Discussion}
\label{sec:approx}
% In this section, we would like to discuss the approximations
% and assumptions introduced in the derivations presented in
% section \ref{sec:theory}.
% 
%\subsection{Introduced approximations and their consequences}
In this article, we have derived expressions for the time-resolved description
of electronic decay processes including nuclear degrees of freedom. In the following,
we would like to discuss the approximations and assumptions
introduced in these derivations presented in section~\ref{sec:theory}.
%we will recapitulate the used approximations and discuss their
%consequences for the simulated spectra.

\begin{enumerate}
 \item We use the Born-Oppenheimer approximation throughout the article. Therefore,
       the derived expressions will not be applicable when strong non-adiabatic
       couplings exist which involve the resonance or the final states.
 \item We assume that the transition dipole moment matrix elements
       $\braket{E|\mu|g}$
       are independent of the energy of the final state.
 \item Such independence of the energy of the final state
       is also assumed for the coupling matrix elements $V_{Er}$.
       Consequently, the effective decay width $W_\lambda$,
       Eq.~(\ref{eq:Wlambda}), is considered to be independent of energy.
 \item In this article, we only consider bound vibrational eigenstates.
       Therefore, the introduced nuclear bases are not necessarily complete.
       This assumption manifests itself in the reduced overall decay widths (Eq.~\eqref{eq:Wlambda})
       and therefore the observed increased lifetimes with respect to
       the purely electronic case. Moreover, these contributions are also missing
       in the simulated spectra. Their inclusion will be subject to future work.
 \item In the above simulations, we discussed time-resolved spectra involving
       multiple resonance states in combination with several final states.
       In a rigorous treatment, it is only possible
       to include either several resonance or several final states,
       but not several states
       of both kinds. In order to still treat their combination,
       we have introduced a few approximations, whose impact we would like discuss
       in more detail.
    \begin{enumerate}
     \item In the derivation of the coefficients of the Fano wavefunction, we assume
           that different resonance states do not couple via the continuum.
           In this case, the only meaningful contribution in Eq.~(\ref{eq:Fano_approx})
           is the one for $\lambda' = \lambda$.
           Phenomenologically, this means
           that no lifetime sharing between the different
           vibrational states of the resonance state is possible.\\
           For the molecular Auger-Meitner process
           with large decay widths, this is in many cases a crude
           approximation. However, for the RICD process with longer
           lifetimes, this approximation becomes feasible, as
           can, e.g., be estimated from the ICD process of the neon
           dimer \cite{Scheit03}.
           %It also means that the principal part
           %of the integral over $E'$ must be zero, which we use later in the derivation.
           %It moreover means that $|V_{E'r}|^2$
           %must be independent of the energy $E'$.
     \item We neglect the impact of $F_{\mu'}(E)$.
           Neglecting $F(E)$ in the purely electronic solution for one resonance
           and one final state results in neglecting a small shift in the resonance
           energy, while the overall shape of the corresponding peak
           remains the same. This effect is more pronounced
           when several resonance states are involved.
           By neglecting $F_{\mu'}(E)$, we neglect
           shifts of all resonance energies, which can
           change the energy differences between them.
           Additionally, we also neglect changes
           in the respective decay widths and hence lifetimes. Strong couplings
           via the continuum will therefore lead to qualitatively
           different spectra than those shown in this article.
           In extreme cases for two resonance states, one resonance state can combine the
           entire decay capability of the system into itself, while the other state
           is stabilized.
    
           It would be beneficial to go beyond the current limitations in future work.
     \item The approximation of negligible couplings of different resonance states
           via the continuum leads to an underdetermined expression in
           Eq.~(\ref{eq:norm_a_sum}). We therefore assume equal contributions of
           $|a_\lambda(E)|^2 W_\lambda (\pi^2 + |z(E)|^2)$ for all vibrational
           resonance states. This approximation is very likely incorrect.
           If the contributions for different $\lambda$ are not equal, it will
           change the relative size of the different contributions and introduce
           additional prefactors for each resonance or indirect contribution. The
           overall shape of the equations would not change though.
    \end{enumerate}
 \item In the Fano wavefunction, the coefficients $b$ of the continuum states
       are expressed as sums of contributions of the resonance states,
       Eq.~(\ref{eq:Fano_coeff_b}).
       In a fashion related to the above assumption,
       we neglect the contributions of all vibrational states $\ket{\chi_{\lambda'}}$
       of the resonance state in this description
       except the one where $\lambda' = \lambda$.
       This greatly reduces the complexity of the expressions
       and facilitates the integration necessary when evaluating the Fano matrix elements, Appendix~\ref{app:Fano_me}.
       The solution of these integrals including the contributions of other states $\ket{\chi_{\lambda'}}$ with $\lambda' \ne \lambda$ will be discussed in a future study.

    \end{enumerate}

%%%%%%%%%%%%%%%%%%%%%%%%%%%%%%%%%%%%%%%%%%%%%%%%%%%%%%%%%%%%%%%%%%%%%%%%%%%%%%%%%
%\input{subsecs/summary}

\section{Summary}
\label{sec:summary}

In this section, we want to summarize the general results
presented in section \ref{sec:cases} within the approximations
discussed in section \ref{sec:approx}.

As we have previously shown in Ref.~\cite{Fasshauer20_1}, the signal shape
for a single decay is
given by the convolution of the corresponding Fano lineshape and
the Fourier transform of the envelope of the exciting laser pulse. If the
mean photon energy of the laser diverges from the resonance energy, the decay
peak is damped, compared to the same peak on resonance (see Figure~4 of Ref.~
\cite{Fasshauer20_1}).

The Fano parameter $q$ (see Eq.~(\ref{eq:Fano_param}))
determines the relative contributions of the different
pathways to the final signal, and hence, the larger $q$ (the larger the relative
contribution of the resonance pathway), the more the maximum is moved towards the
kinetic energy corresponding to the resonance and the minimum of the
characteristic Fano lineshape is moved to $-\infty$
(see Eqs.~(\ref{eq:Fano_max_el}, \ref{eq:Fano_min_el})).
Hence, the lineshape becomes a Lorentzian convoluted with a Gaussian,
a so-called Voigt profile.
Furthermore, the smaller the decay width (the longer the lifetime), the less
pronounced the Fano lineshape is, according to Eq.~(\ref{eq:Fano_param}).

Specific to several possible final states in general, and vibrational final
states in particular, are several direct pathways and, correspondingly, several
peaks at kinetic energies lower than the decay between the vibrational ground
states. This can clearly be seen in Figure~\ref{fig:1lxmE}, where the spectrum
after \qty{200}{fs} is compared between cases 1, 2 and 3.

%\begin{figure}[h]
%\centering
% \includegraphics[width=\columnwidth]{pics/energy_1lxm.pdf}
%\caption{}
%\label{fig:1lxmE}
%\end{figure}

\begin{figure}[h]
\centering
 \includegraphics[width=\columnwidth]{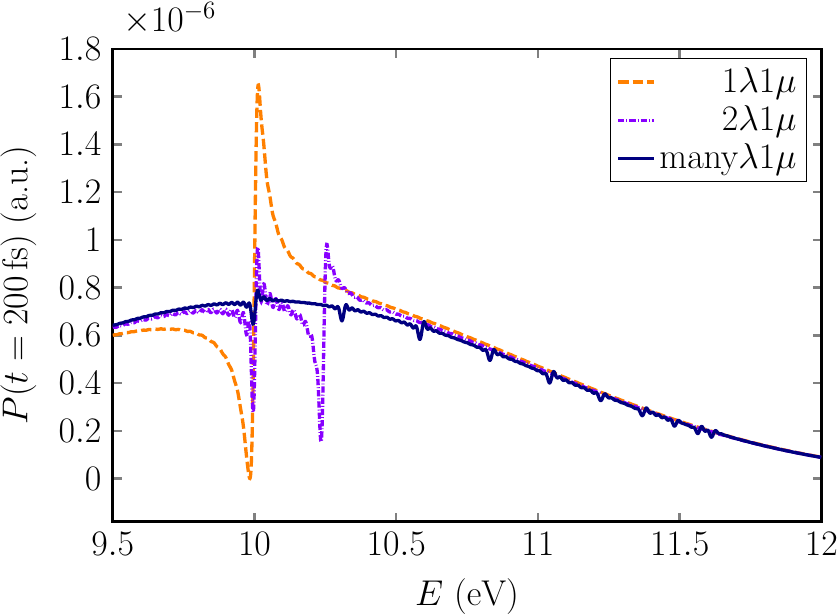}
 \caption{(Colour online) Spectra at time $t=\qty{200}{fs}$
    for cases 1 (dashed curve in orange) and 4.1 (dash-dotted curve in violet)
    as well as a case with ten vibrational resonance states (solid curve in dark blue).}
\label{fig:xl1mE}
\end{figure}

Analogously, we compare the spectra \qty{200}{fs} after the centre of the pulse
for several vibrational resonance states in Figure~\ref{fig:xl1mE}. In this case,
the kinetic energy increases with the vibrational quantum number of the decaying state.

When several vibrational resonance states exist, all but one are automatically
excited off-resonance, and hence, their resonance parts are weighted differently by the
corresponding direct pathway, which
can clearly be seen in Figure~\ref{fig:xl1mE}.

The Fano lineshape is created by the interference between the contributions
of the different pathways to the total wavefunction.
As we have seen in all cases, this lineshape is affected by the inclusion
of the nuclear degrees of freedom. The Franck-Condon overlaps alter the relative
contributions of the different pathways and hence lead to different effective
Fano parameters.% $q_\text{eff}$ (see Eq.~\eqref{eq:eff_Fano_param}).

We have earlier shown in
Ref.~\cite{Fasshauer20_1} that the interference between the different pathways
is also responsible for the
predicted oscillations in time and energy even for only one resonance and one
final state.
Inspection of Eq.~(\ref{eq:ampl_working_eq}) as well as
Eqs.~(\ref{eq:dir_ampl}) and (\ref{eq:res_ampl}) infers that each vibrational resonance
state contributes separately to the wavefunction. These contributions can therefore
also interfere and cause additional time-dependent variations of the signals, which
follow Eq.~(\ref{eq:res_res_int}). The corresponding interference pattern can then be used
to gather information about the wavepacket of the system and therefore the electronic
and nuclear motion.

The resonance-resonance interference signal is only part of the
overall signal.
In order to observe it, one needs
a system with a large $q$ and to focus on energy regions with sufficient intensity
but not at the kinetic energies corresponding to the
resonance energies themselves. As discussed in case 4,
a short initiating pulse can help
creating the necessary peak intensity away from the kinetic energies of the electron corresponding
to a resonant excitation.

The complexity of a spectrum as well as its time dependency
is increased by the number of involved decaying and
final states and so is the interference spectrum. In addition to the different
interferences, overlapping signals can also lead to patterns which merely look
similar to interference patterns.
A Fourier analysis, though, can help to distinguish between overlapping
signals and interference patterns, as we have discussed in detail
in sections \ref{sec:1l2m} and \ref{sec:2l1m}.

%%%%%%%%%%%%%%%%%%%%%%%%%%%%%%%%%%%%%%%%%%%%%%%%%%%%%%%%%%%%%%%%%%%%%%%%%%%%%%%%%
%\input{subsecs/ack}

\section{Acknowledgements}
E.\ F.\ and A.\ V.\ R.\ gratefully acknowledge funding through the
Core Facility LISA$^+$ of the University of Tübingen.
E.\ F.\ acknowledges the COST Action CA18222 (Attosecond Chemistry).
We would further like to
thank Nicolas Sisourat for a helpful comment on the completeness of our
basis of vibrational states.

%%%%%%%%%%%%%%%%%%%%%%%%%%%%%%%%%%%%%%%%%%%%%%%%%%%%%%%%%%%%%%%%%%%%%%%%%%%%%%%%%
%\input{subsecs/appendix}

\appendix

\section{Evaluation of the Fano matrix elements}
\label{app:Fano_me}
The Fano integrals can now be solved by contour integration in the negative
complex half-plane, where we employ Eq.~(\ref{eq:Fano_projection}), use
$\braket{r|r}=1$, $\braket{E|E'}=\delta(E-E')$ and $\braket{E|r}=0$
as well as $\braket{\chi_\lambda|\chi_{\lambda'}} = \delta_{\lambda,\lambda'}$
and $\braket{\chi_\mu|\chi_{\mu'}} = \delta_{\mu,\mu'}$
and furthermore ignore the effects of $F_{\mu'}(E)$:

\begin{widetext}
\begin{align}
 &% \sum\limits_{\lambda,\lambda'}
    \bra{\chi_{\lambda}} \braket{r|U_\mathrm{F}(t'',t')| r} \ket{\chi_{\lambda'}} \nonumber\\
 %=& \sum\limits_{\lambda,\lambda',\lambda''}
 =& \sum\limits_{\lambda''}
 \int\mathrm{d}\underline{E}
    \bra{\chi_\lambda} \bra{r} a_{\lambda''}(\underline{E})
     \ket{r} \ket{\chi_{\lambda''}}
    \bra{\chi_{\lambda''}} \bra{r} a^*_{\lambda''}(\underline{E})
     \ket{r} \ket{\chi_{\lambda'}}  \exp[-i\underline{E} (t''-t')] \\
% =& \sum\limits_{\lambda,\lambda',\lambda''}
 =& \sum\limits_{\lambda''}
 \int\mathrm{d}\underline{E}
     \,\, \delta_{\lambda,\lambda''} \delta_{\lambda',\lambda''} \,\,
     a_{\lambda''}(\underline{E})
     a^*_{\lambda''}(\underline{E})
     \exp[-i\underline{E} (t''-t')] \\
 =&% \sum\limits_{\lambda,\lambda'}
 \int\mathrm{d}\underline{E} \,\,
    \exp[-i\underline{E} (t''-t') ] \, |a_\lambda(\underline{E})|^2
    \delta_{\lambda,\lambda'}\\
 =&% \sum\limits_{\lambda,\lambda'}
 \int\mathrm{d}\underline{E}
     \frac{\exp[-i\underline{E} (t''-t')]  W_\lambda }
     {N_\lambda ((\underline{E} - E_r - E_\lambda)^2 + \pi^2 W_\lambda^2)} 
     \delta_{\lambda,\lambda'}
\end{align}

Note that the effective decay width $W_\lambda$ is,
according to Eq.~(\ref{eq:Wlambda}),
in principle dependent on the energy of the final state,
$W_\lambda = \sum_{\mu'} |V_{(\underline{E}-E_{\mu'})r}|^2 \,
| \braket{\chi_{\lambda}|\chi_{\mu'}} |^2 = W_{\lambda}(\underline{E})$.
We now assume the electronic coupling matrix element $V_{Er}$
to be independent of the continuous kinetic energy of the emitted
electron and to be sufficiently described by the discretized
energy difference between the resonance and the final state
and therefore denote it $V_r$.
We incorporate the discretized final states by summing over them
in the expression for the effective decay width,
which now has become independent of the energy:
$W_{\lambda}(\underline{E}) = |V_r|^2 \sum_{\mu'}
| \braket{\chi_{\lambda}|\chi_{\mu'}} |^2 = W_\lambda$.
Treating it as a constant inside the integral and evaluating the latter yields the Fano matrix element:

\begin{align}
% =& V_r^2 \int\mathrm{d}\underline{E} \,\, \frac{ \exp[-i\underline{E} (t''-t')] }
%                                             {(\underline{E} - E_r)^2 + \pi^2 V_r^4} \\
% =& V_r^2 \int\mathrm{d}\underline{E} \,\, \frac{ \exp[-i\underline{E} (t''-t')] }
%           {(\underline{E} - E_r + i\pi V_r^2) \, (\underline{E} - E_r - i\pi V_r^2)} \\
% =& V_r^2 \, \exp[ -iE_r (t''-t') ]
%    \int\mathrm{d}\tilde{E} \,\, \frac{ \exp[-i\tilde{E} (t''-t')] }
%           {(\tilde{E} + i\pi V_r^2) \, (\tilde{E} - i\pi V_r^2)} \\
% =& 2\pi i \, V_r^2 \, \exp[ -iE_r (t''-t') ] \, a_{-1} \\
 \frac{1}{N_\lambda} %\sum\limits_{\lambda,\lambda'}
 \delta_{\lambda,\lambda'}
 \exp[ -i (E_r + E_\lambda -i \pi W_\lambda) (t'' - t')]
\end{align}
%\par\noindent\rule{5cm}{0.4pt}
\hrule height 0.4pt depth 0pt width \linewidth \relax
%\end{widetext}

%%%%%%%%%%%%%%%%%%%%%%%%%%%%%%%%%%%%%%

%\begin{widetext}
\begin{align}
 &% \sum\limits_{\lambda,\mu''}
 \int \mathrm{d}E''
   \bra{\chi_\lambda} \braket{r|U_\mathrm{F}(t'',t')| E''} \ket{\chi_{\mu''}} \nonumber \\
% =& \sum\limits_{\lambda,\lambda'',\mu'',\tilde{\mu}}
 =& \sum\limits_{\lambda'',\tilde{\mu}}
 \int \mathrm{d}E''
    \int\mathrm{d}\underline{E} \int\mathrm{d}\tilde{E} \,\,
    \exp[-i\underline{E} (t''-t')] \,
    \bra{\chi_\lambda} \bra{r} a_{\lambda''}(\underline{E}) \ket{r} \ket{\chi_{\lambda''}}
    \bra{\chi_{\tilde{\mu}}} \bra{\tilde{E}} b_{\tilde{E},\tilde{\mu}}^*(\underline{E})
    \ket{E''} \ket{\chi_{\mu''}} \\
 =& %\sum\limits_{\lambda,\mu''}
 \int \mathrm{d}E''
    \int\mathrm{d}\underline{E} \,\,
    \exp[-i\underline{E} (t''-t')] \,
    a_\lambda(\underline{E})
    b_{E'',\mu''}^*(\underline{E})      \label{eq:Fano_element_cont_with_b}
    \\
% =& V_r^2 \int\mathrm{d}\underline{E} \,\, \frac{ (\underline{E} - E_r)
%                                                 \exp[-i\underline{E} (t''-t')] }
%                                             {(\underline{E} - E_r)^2 + \pi^2 V_r^4} \\
% =& V_r^2 \int\mathrm{d}\underline{E} \,\, \frac{ (\underline{E} - E_r)
%                                                  \exp[-i\underline{E} (t''-t')] }
%           {(\underline{E} - E_r + i\pi V_r^2) \, (\underline{E} - E_r - i\pi V_r^2)} \\
% =& V_r^2 \, \exp[ -iE_r (t''-t') ] 
%           \int\mathrm{d}\tilde{E} \,\, \frac{ \tilde{E} \exp[-i\tilde{E} (t''-t')] }
%           {(\tilde{E} + i\pi V_r^2) \, (\tilde{E} - i\pi V_r^2)} \\
% =& 2\pi i \, V_r^2 \, \exp[ -iE_r (t''-t') ] \, a_{-1} \\
\begin{split}
			=& \int \mathrm{d}E''
			\int\mathrm{d}\underline{E} \,\,
			\exp[-i\underline{E} (t''-t')] \,
			a_\lambda(\underline{E}) \,
			\sum\limits_{\lambda'}\,\,
			\Bigl(
			\frac{V_{rE''} \, a^*_{\lambda'}(\underline{E}) \braket{\chi_{\lambda'} | \chi_{\mu''}}}
			{\underline{E} - E'' - E_{\mu''}} \\
			&+ \, 
			\frac{\underline{E}-E_r - E_{\lambda'}}
			{W_{\lambda'}} \,
			V_{rE''} \, a^*_{\lambda'}(\underline{E})
			\delta(\underline{E} - E'' - E_{\mu''}) 
			\braket{\chi_{\lambda'} | \chi_{\mu''}}
			\Bigr)
\end{split} \\
\begin{split}
			=& \int\mathrm{d}\underline{E} \,\,
			\exp[-i\underline{E} (t''-t')] \,
			a_\lambda(\underline{E}) \,
			\sum\limits_{\lambda'}\,\,
			\Bigl(
			a^*_{\lambda'}(\underline{E})
			\braket{\chi_{\lambda'} | \chi_{\mu''}} \,
			\mathcal{P}\mspace{-5mu}\int\mathrm{d}E''
			\frac{V_{rE''}}
			{\underline{E} - E'' - E_{\mu''}} \\
			&+ \, \frac{\underline{E}-E_r - E_{\lambda'}}
			{W_{\lambda'}} \,
			V_{r(\underline{E} - E_{\mu''})} a^*_{\lambda'}(\underline{E})
			\braket{\chi_{\lambda'} | \chi_{\mu''}}
			\Bigr)
		\end{split}
\end{align}

We again assume the coupling matrix element $V_{Er}$
to be independent of the final state, $V_{Er} = V_r$ and $V_{rE} = V^*_r$.
Then the principal value of the integral in the first summand is zero.
Furthermore, the effective decay width again becomes independent
of the energy, $W_{\lambda}(\underline{E}) = W_\lambda$.
The Fano matrix element then reads as
\begin{align}
		& \int\mathrm{d}\underline{E} \,\,
		\exp[-i\underline{E} (t''-t')] \, V^*_r \,
		a_\lambda(\underline{E}) \,
		\sum\limits_{\lambda'}\,\,
		\frac{\underline{E}-E_r - E_{\lambda'}}
		{W_{\lambda'}} \,
		a^*_{\lambda'}(\underline{E})
		\braket{\chi_{\lambda'} | \chi_{\mu''}}   \, .
\end{align}

Due to the sum over $\lambda'$, several terms contribute to the Fano matrix element. This renders further equations quite involved.
Therefore, we make the assumption that the only vibrational eigenstate $\ket{\chi_{\lambda'}}$ of the electronic resonance state $\ket{r}$
that contributes to the description of the chosen final state $\ket{E''}\ket{\chi_{\mu''}}$ is the chosen vibrational eigenstate $\ket{\chi_\lambda}$.
This means that in Eq.~(\ref{eq:Fano_coeff_b}) for the Fano coefficient of the final state, $\delta_{\lambda,\lambda'}$ is introduced
so that all $\lambda'$ are replaced by $\lambda$ and the sum over all $\lambda'$ drops out.
Since the present expression for the Fano matrix element
has been derived by inserting Eq.~(\ref{eq:Fano_coeff_b})
into Eq.~(\ref{eq:Fano_element_cont_with_b}), the same is true here.

\begin{align}
    & \int\mathrm{d}\underline{E} \,\,
    \exp[-i\underline{E} (t''-t')] \, V^*_r \,
    | a_\lambda(\underline{E}) |^2 \,
    \frac{\underline{E}-E_r - E_\lambda}
    {W_\lambda} \,
    \braket{\chi_\lambda | \chi_{\mu''}}    \\
    =& \, \frac{1}{N_\lambda} V^*_r
    \braket{\chi_{\lambda} | \chi_{\mu''}}  \,
    \int\mathrm{d}\underline{E} \,\,
    \frac{(\underline{E}-E_r - E_\lambda) \,
    \exp[-i \underline{E} (t''-t')]}
    {(\underline{E}-E_r - E_\lambda)^2 + \pi^2 W_\lambda^2}    \\
    =& -\frac{i \pi}{N_\lambda} V^*_r
    \braket{ \chi_{\lambda} | \chi_{\mu''} }    \,
    \exp[ -i (E_r + E_\lambda -i \pi W_\lambda) (t'' - t')]
\end{align}

% \begin{align}
%      =& -\frac{i \pi}{N_\lambda}
%      %\sum\limits_{\lambda,\mu''}
%      V_{r(E_r+E_\lambda-E_{\mu''})} 
%      \braket{ \chi_{\lambda} | \chi_{\mu''} }
%      \exp[ -i (E_r + E_\lambda -i \pi W_\lambda) (t'' - t')]
% \end{align}
\end{widetext}

%\bibliographystyle{apsrev4-1}
%\bibliography{theolit}


\end{document}